\documentclass[12pt]{article}

\usepackage{scicite}
\usepackage{epsf}

\usepackage{ulem}
\usepackage{color}

\usepackage[squaren,Gray]{SIunits}
\usepackage{times}
\usepackage{amsmath}
\usepackage{amsfonts}
\usepackage{amssymb}
\usepackage{graphicx}
\usepackage{color}
\usepackage{hyperref}
\usepackage{chngcntr}

\usepackage[utf8]{inputenc}
\usepackage[english]{babel}
\usepackage[table,xcdraw]{xcolor}
\usepackage{amsmath,amsfonts,amssymb}
\usepackage{geometry}
\usepackage{graphicx}
\usepackage{color}
\usepackage{xcolor}
\usepackage{empheq}
\usepackage{hyperref}
\usepackage{bm}
\usepackage{url}
\usepackage{diagbox}
\usepackage{array}
\usepackage{setspace}
\usepackage{multirow}

\newcommand*{\flav}[1]{\textcolor{black}{#1}}

\newcommand*{\purple}[1]{\textcolor{violet}{#1}}

\newcommand*{\alex}[1]{\textcolor{black}{#1}}

\newcommand{\inm}{\textrm{in}}
\newcommand{\outm}{\textrm{out}}

\newcommand{\ugg}{\mathbf{u}}

\newcommand{\uout}{\ugg_\outm}
\newcommand{\rin}{\mathbf{r_\inm}}
\newcommand{\rout}{\mathbf{r_\outm}}
\newcommand{\rg}{\mathbf{r}}
\newcommand{\rhoin}{\boldsymbol{\rho}_\inm}
\newcommand{\rhoout}{\boldsymbol{\rho}_\outm}

\newcommand{\vg}{\mathbf{v}}

\newcommand{\rhog}{\boldsymbol{\rho}}

\newcommand{\R}{\mathbf{R}}

\newenvironment{sciabstract}{%
\begin{quote} \bf}
{\end{quote}}
\renewcommand{\thesubsection}{\Roman{subsection}.}

\title{Ultrasound matrix imaging for \alex{3D} transcranial in-vivo localization microscopy} 

\author
{Flavien Bureau${}^{1,\dag}$, Louise Denis${}^{2,\dag}$,Antoine Coudert${}^{2}$, Mathias Fink${}^{1}$,\\ Olivier Couture${}^{2}$, and Alexandre Aubry${}^{1,\ast}$\\
\normalsize{$^1$ PSL University, ESPCI Paris, CNRS, Institut Langevin, Paris, France}
\normalsize{}\\
\normalsize{$^2$Sorbonne Universit\'{e}, CNRS, INSERM, Laboratoire d'Imagerie Biomedicale, Paris, France}\\
\normalsize{$^\dag$These authors contributed equally to this work.}\\
\normalsize{$^\ast$To whom correspondence should be addressed; E-mail:  alexandre.aubry@espci.fr.}
\\
\normalsize{{\textbf{Summary Sentence}: Ultrasound matrix localization microscopy provides}}\\
\normalsize{{a three-dimensional, artifacts-free and super-resolved image of brain vessels.}}
}

\date{}
\begin{document}
\baselineskip24pt
\maketitle
\begin{sciabstract}
Transcranial ultrasound imaging is usually limited by skull-induced attenuation and high-order aberrations. By using contrast agents such as microbubbles in combination with ultrafast imaging, not only can the signal-to-noise ratio be improved, but super-resolution images down to the micrometer scale of the brain vessels can be obtained. However, ultrasound localization microscopy (ULM) remains impacted by wave-front distortions that limit the microbubble detection rate and hamper their localization. In this work, we show how \flav{ultrasound matrix imaging (UMI)}, which relies on the prior recording of the reflection matrix, can provide a solution to those fundamental issues. As an experimental proof-of-concept, an in-vivo reconstruction of deep brain microvessels is performed on three anesthetized sheeps. The compensation of wave distortions is shown to drastically enhance the contrast and resolution of ULM.  This experimental study thus opens up promising perspectives for a transcranial and non-ionizing observation of human cerebral microvascular pathologies, such as stroke.
\end{sciabstract}

\section*{INTRODUCTION}

Transcranial imaging is essential to understand the complex vascular mechanisms underlying pathologies. For cerebrovascular events, such as strokes, X-rays and computed tomography (CT) are currently used to observe hemorrhagic events~\cite{Bahrami} and magnetic resonance imaging (MRI) to diagnose and date ischemic stroke~\cite{Scheldeman2020}. Although these methods allow visualization of many neurovascular diseases, MRI or CT-scans require large and expensive equipment that is only primarily available in health care centers.

Alternatively, transcranial ultrasound is used {in 
Doppler mode} to monitor reperfusion after a stroke ~\cite{kirsch_advances_2013}. This technique allows portable real-time imaging at low cost. 
{However, it} 
remains a major challenge as the complexity of the skull layer with its unpredictable porosity causes severe attenuation, strong aberrations {and} multiple scattering, leading to a drastic reduction in resolution and 
 contrast {of the ultrasound image}.

Microbubbles (MB) are often used to improve the {visualization of blood flow in ultrasound imaging}
~\cite{versluis_ultrasound_2020,tarighatnia_recent_2022,liu_combining_2014}. In the clinic, they are injected intravenously and are excellent contrast agents due to their high impedance mismatch compared to soft tissue{s}. 
The use of ultrafast ultrasound scanners, i.e. with high {frame rates}, has also enabled organ perfusion dynamics to be studied with greater precision~\cite{couture_ultrafast_2009, couture_ultrasound_2012}.

In addition to enhancing the blood signal, microbubbles can also be used to generate super-resolved images. This is the principle of ultrasound localization microscopy (ULM)~\cite{errico_ultrafast_2015, couture_ultrasound_2018, christensen-jeffries_super-resolution_2020, song_super-resolution_2023}, which is based on a similar idea developed a few years earlier in fluorescence microscopy~\cite{betzig_imaging_2006,rust_stochastic_2006}. By separating the echoes of individual injectable microbubbles, their position can be localized with a sub-wavelength precision and tracked over time.
This allows an accuracy that no longer depends on the diffraction limit but on the accuracy of detection of their center. As a consequence, ULM results in a significant improvement in resolution by a factor of ten~\cite{desailly_resolution_2015} compared to standard ultrasound images. ULM has already proved its worth in many organs in depth, both pre-clinically~\cite{errico_ultrafast_2015, hingot_subwavelength_2017, lowerison_aging-related_2022, claron_large-scale_2021, demeulenaere_coronary_2022, hansen_robust_2016, taghavi_vivo_2021, zhu_3d_2019, lin_3-d_2017} and clinically~\cite{opacic_motion_2018, huang_super-resolution_2021, demene_transcranial_2021, Denis2023, bodard_ultrasound_2023, bodard_visualization_2024, knieling_2023, yan_transthoracic_2024,denis_transcranial_2024}. The real-time information contained in the displacement of microbubbles enables complex vascular networks to be mapped at the micrometer scale, even going as far as the capillary-level reconstruction of functional units, such as glomeruli in the kidney (sULM)~\cite{Denis2023,bodard_ultrasound_2023,bodard_visualization_2024}. ULM imaging has even been successfully used in transcranial clinical imaging of the adult human brain~\cite{demene_transcranial_2021}, where aberration correction is critical~\cite{soulioti_super-resolution_2020, robin_vivo_2023}.

Recently, 3D ULM\cite{heiles_ultrafast_2019, chavignon_3d_2022, demeulenaere_coronary_2022, demeulenaere_vivo_2022, mccall_longitudinal_2023, bourquin_quantitative_2024, lei_vivo_2023, wei_high-frame-rate_2023, coudert_3d_2024, xing_towards_2024, chabouh_whole_2024}  has overcome many 2D limitations, such as the projection of 3D structures in 2D, suboptimal tracking of microbubbles, and out-of-plane probe movements that prevent vascular reconstruction. Nevertheless, aberrations caused by the skull remain a problem for ULM, especially in 3D. The sensitivity of bubble detection is lower, which affects the contrast of the super-resolved image, but also the accuracy of bubble localization, where the high sidelobes of a degraded PSF can cause artifacts such as vessel doubling~\cite{xing_phase_2023, robin_vivo_2023}.

To solve the fundamental problem of aberrations, adaptive focusing was introduced years ago in astronomy to compensate for phase distortions undergone by starlight
when it
passes through the atmosphere \cite{babcock_possibility_1953, labeyrie_attainment_1970, muller_real-time_1974, roddier_adaptive_1999}. {This idea was then transposed to ultrasound imaging by tailoring a focusing law that either optimizes an image metrics~\cite{hirama_adaptive_1982, nock_phase_1989} or maximizes the spatial correlations of back-scattered echoes}\cite{flax_phase-aberration_1988, mallart_adaptive_1994, montaldo_time_2011, masoy_iteration_2005, odonnell_phase-aberration_1988}. 
In the specific case of ULM, several methods have been proposed to correct the aberrations caused by the skull. A first strategy~\cite{xing_phase_2023} is based on deep learning algorithms trained with numerical simulations. Although it gives impressive in-vivo results for small animals, it is hampered by the specificity of the training set, so its transfer to other {experimental configurations} (probe, organ, SNR) may require tedious training procedures. The second method~\cite{robin_vivo_2023} uses microbubbles as ultrasound guide stars but suffers several drawbacks. On the one hand, microbubbles should be detected and isolated enough, which can be extremely difficult to assess because of the strongly distorted focal spot across the skull. On the other hand, the number of detected micro-bubbles should be large enough for self-averaging in each isoplanatic patch. \alex{Last but not least, the demonstration was limited to 2D imaging using a 1D phased array probe, while skull heterogeneities are 3D distributed and require a bi-dimensional control of the ultrasonic wave-field to properly compensate for such high-order aberrations~\cite{bureau_three-dimensional_2023}.}

Interestingly, inspired by previous seminal works~\cite{varslot_eigenfunction_2004,robert_greens_2008,badon_distortion_2020}, a matrix approach of ultrasound imaging has been recently developed to overcome aberrations without any guide star~\cite{lambert_reflection_2020,lambert_distortion_2020}. Based on the recording of the reflection associated with the ultrasonic probe and the linearity of the wave equation, it allows the simulation of wave focusing in post-processing and an iteration of this process towards local adaptive focusing laws in ultrasonic speckle~\cite{lambert_ultrasound_2022b}. An optimized contrast and close-to-ideal resolution is then obtained for each voxel of the ultrasound image. \flav{Ultrasound Matrix Imaging (UMI)} has been successfully applied to transcranial 3D imaging of a head phantom~\cite{bureau_three-dimensional_2023}.

The goal of this paper is to demonstrate the interest of UMI for in-vivo \alex{3D} brain imaging. In particular, we will show how: (i) UMI can be used to quantify aberrations and multiple scattering in in-vivo transcranial imaging; (ii) UMI can exploit static speckle from brain tissue to tailor complex \alex{2D} focusing laws \alex{in a strong aberration and multiple scattering regime}; (iii) \alex{ULM} can leverage these focusing laws \alex{despite motion} to increase the detection rate of microbubbles and improve their localization; (iv) UMI can be fruitfully combined to ULM in order to provide a three-dimensional, artifacts-free, super-resolved and contrasted image of brain vessels.

To that aim, an experimental proof-of-concept will consist in brain imaging on three anesthetized sheeps. The first step of the experimental procedure consists in the acquisition of the reflection matrix and the estimation of local focusing laws by means of UMI. The second step consists in a microbubble injection and the beamforming of ultra-fast images using the focusing laws provided by UMI. The last part consists in the usual ULM post-processing that consists in localizing and tracking micro-bubbles. Final ULM image contrast will be shown to be drastically improved by the better detection rate of smaller micrometer-sized vessels provided by UMI. The comparison with MRI will show how UMI allows the removal of artifacts from which the initial ULM image suffered.

\section*{RESULTS}
\begin{figure}[ht]
\includegraphics[width=\textwidth]{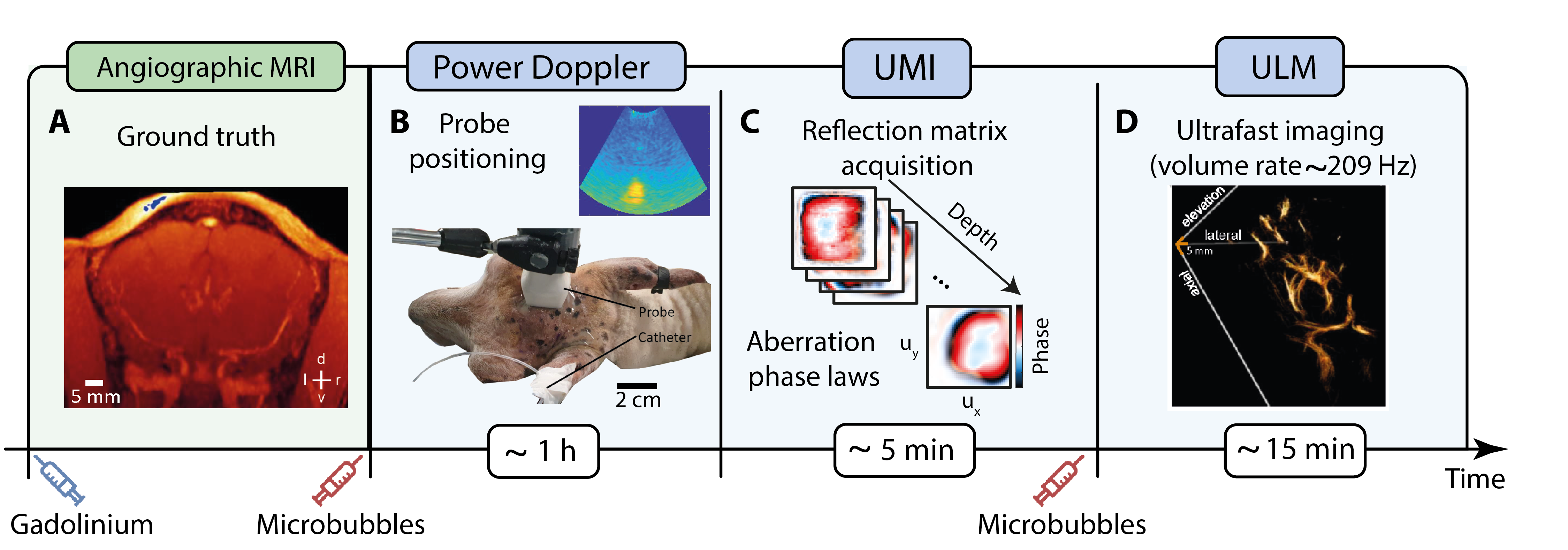}
    \caption{\textbf{Sheep experiment : A multi-sequence acquisition}. (\textbf{A}) Angiographic MRI. (\textbf{B}) Transcranial power Doppler for positioning the probe.  (\textbf{C}) Acquisition of the reflection matrix to estimate local aberration laws with 3D-UMI.  (\textbf{D}) Ultrafast imaging for super-resolved ULM images. The images shown in this figure are for illustrative purposes only and do not provide quantitative results.}
    \label{fig1_acquisition}
\end{figure}

\subsection*{Reflection matrix in a virtual source basis}

{In this paper, we demonstrate transcranial  brain imaging by overcoming the major challenges of: (\textit{i}) skull bone aberration with 3D UMI; (\textit{ii}) diffraction-limited resolution with the super-resolution feature provided by ULM. As a proof-of-concept, the same experimental protocol (Fig.~\ref{fig1_acquisition}) has been applied to three anesthetized sheeps (Methods). A $32\times 32$ array of transducers is placed on the shaved head of each sheep directly over the crest of the frontal bone (Fig.~\ref{fig1_acquisition}B). The probe is driven in 1-2 frequency bandwidth in order to limit the attenuation by skull bone~\cite{fry_acoustical_1978}. The first ultrasound sequence consists in recording a high-dimension reflection matrix $\mathbf{R}_{\mathbf{uv}}(\tau)$ (Fig.~\ref{fig1_acquisition}C) that contains the scattered wave-field $R(\mathbf{u}_\textrm{out},\mathbf{v}_\textrm{in},\tau)$ recorded by each transducer $\mathbf{u}_\textrm{out}$ of the probe as a function of the echo time $\tau$ for a set of 324 incident diverging waves. This illumination basis is associated with a set of virtual transducers $\mathbf{v}_\textrm{in}$ placed behind the probe (see Methods and Fig.~\ref{fig_supp6_virtual_sources}). It maximizes the signal-to-noise ratio across the targeted field-of-view~\cite{provost_3d_2014}. \alex{The number of incident waves is dictated by our choice of the maximal angular range spanning from $-30^{\textrm{o}}$ to $+30^{\textrm{o}}$ (Methods).}}\\
\begin{figure}[ht]
\includegraphics[width=\textwidth]{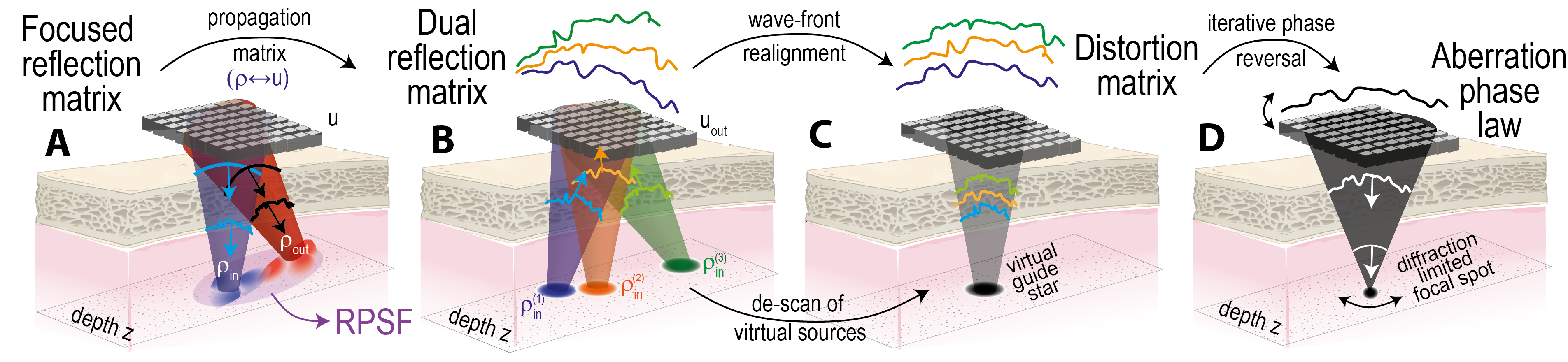}
    \caption{\textbf{Principle of ultrasound matrix imaging}. (\textbf{A}) Focused reflection matrix contains the time-gated response between virtual source ($\bm{\rho}_{\textrm{in}}$) and detector ($\bm{\rho}_{\textrm{out}}$) located at the same depth $z$. (\textbf{B}) An output projection of the focused reflection matrix in the transducer basis ($\mathbf{u}_{\textrm{out}}$) provides the reflected wave-fronts induced by each virtual source ($\bm{\rho}_{\textrm{in}}$) at depth $z$. (\textbf{C}) Those wave-fronts are realigned to extract the wave-front distortions induced by the mismatch between the real speed-of-sound distribution and the wave velocity model. Seen from the focused basis, this operation leads to an angular de-scan of each virtual source at the same position, leading to the synthesis of a virtual guide star. (\textbf{D}) Exploiting the correlations between each distorted wave-front, an iterative phase reversal algorithm extract an aberration phase law that can be used to compensation for wave distortions induce by the skull and ideally retrieve a diffraction-limited focal spot across the field-of-view.}
    \label{fig1_UMI}
\end{figure}

\subsection*{Quantifying skull-induced aberrations and scattering}

\begin{figure}[ht]
\centering
\includegraphics[width=\textwidth]{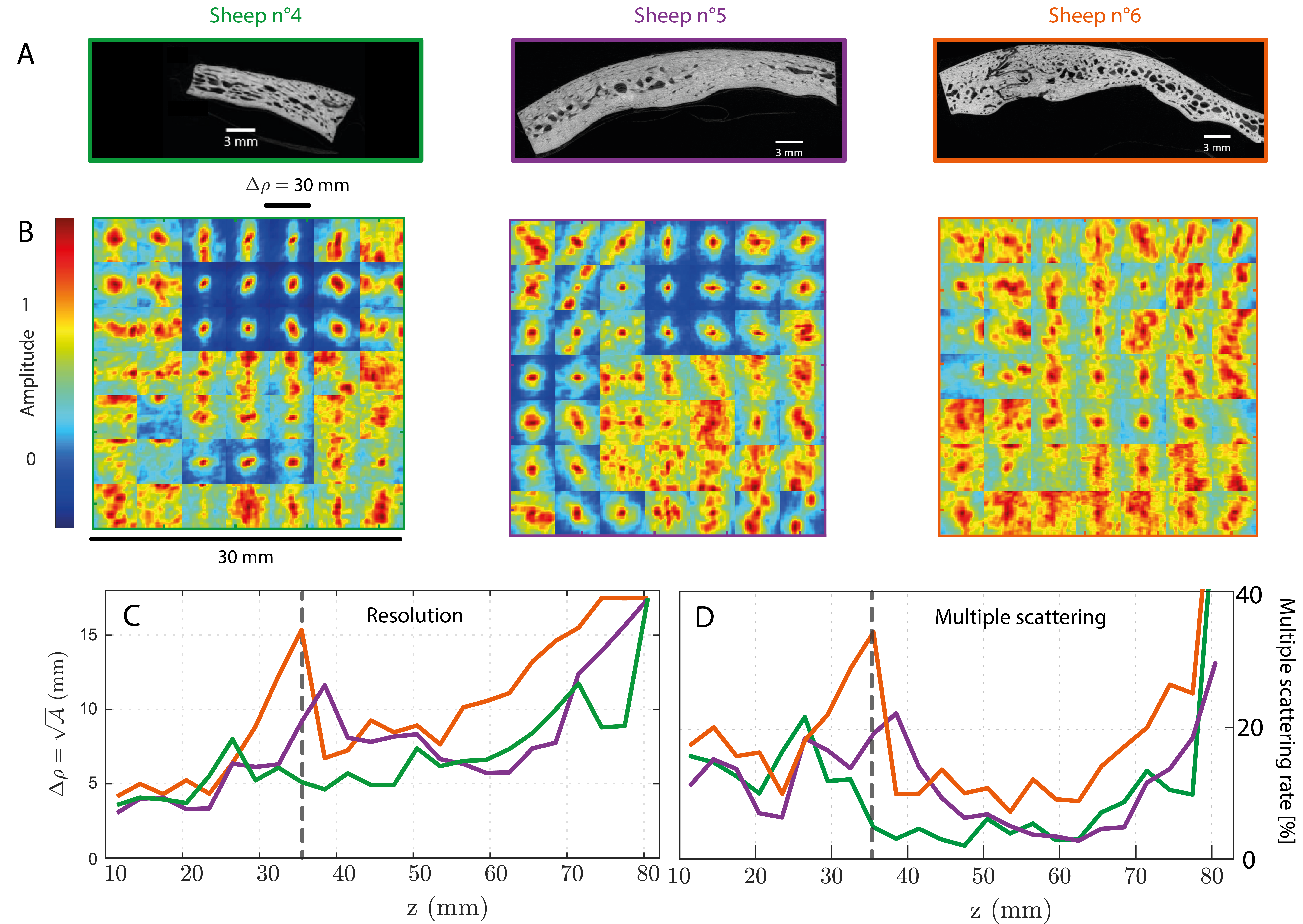}
\caption{ \flav{\textbf{Influence of cranial heterogeneities on aberration and multiple scattering.}  (\textbf{A}) Micro-CT of each sheep skull. Left: sheep n$^{\textrm{o}}$4, middle: sheep n$^{\textrm{o}}$5, right: sheep \textbf{6}.  (\textbf{B}) Corresponding RPSF maps at depth $z=35$ mm. (\textbf{C} and \textbf{D}) RPSF extension and multiple scattering rate as a function of depth, extracted by UMI, for sheeps n$^{\textrm{o}}$4 (green), 5 (purple line) and 6 (orange line).}}
\label{fig7_skull_microct}
\end{figure}

In contrast with standard beamforming that relies on the principle of confocal imaging, the cornerstone of UMI consists in decoupling the input and output focal spots in post-processing. Mathematically, such a beamforming procedure can be written as follows:
\begin{equation}
R(\bm{\rho}_\textrm{out},\bm{\rho}_{\textrm{in}},z)=\sum_{\mathbf{v}_\inm}\sum_{\uout}R\left(\uout,\mathbf{v}_\inm,\tau_\textrm{in}(\mathbf{v}_\textrm{in},\bm{\rho}_{\textrm{in}},z)+\tau_{\textrm{out}}(\uout,\bm{\rho}_{\textrm{out}},z)\right),
\label{generalbemaformchap5}
\end{equation}
where $\rin=(\rhoin,z)$ and $\rout=(\rhoout,z)$ are the coordinates of the input and output focal points, respectively. 
$\tau_{\textrm{in}}$ and $\tau_{\textrm{out}}$ are the expected travel times for the wave to go from the source/detector to the input/output focusing point (Methods, Eq.~\ref{travel_time}). The result of Eq.~\ref{generalbemaformchap5} is a focused reflection matrix $\mathbf{R}_{\bm{\rho \rho}}(z)$ whose coefficients $R(\rhoout,\rhoin,z)$ probe the cross-talk between a virtual source and detector at lateral positions $\rhoin$ and $\rhoout$, respectively, both located at the same depth $z$ inside the medium  (Fig.~\ref{fig1_UMI}A). 
This matrix enables a local quantification of the focusing quality and multiple scattering by providing a local point spread function in reflection (RPSF)~\cite{lambert_distortion_2020,lambert_ultrasound_2022a,bureau_three-dimensional_2023} (Methods, Eq.~\ref{RPSF}). Figure~\ref{fig7_skull_microct}\flav{B display maps of RPSFs at depth $z=35$ mm for each sheep}. In the single scattering regime and in absence of aberrations, the RPSF would exhibit a confocal diffraction-limited peak. On the contrary, the various RPSFs here display a distorted focal spot induced by wave velocity heterogeneities on top of an incoherent background due to multiple scattering events induced by the skull. Strikingly, whatever the imaging depth, the RPSF looks more distorted and less contrasted for sheep n$^{\textrm{o}}$6 than for sheep n$^{\textrm{o}}$\flav{4 and 5} (Fig.~\ref{fig7_skull_microct}B). 

On the one hand, the aberration level can be quantified by investigating the spatial extent $\delta \rho$ of the RPSF ~\cite{lambert_reflection_2020,lambert_ultrasound_2022b} (Fig.~\ref{fig_supp3_resolution}). The depth evolution of $\delta \rho$ is shown in Fig.~\ref{fig7_skull_microct}\flav{C}. A general trend is a degradation of the resolution with depth as predicted by diffraction theory but also a more important aberration level in sheep n$^{\textrm{o}}$6. This difference of behavior can be understood by looking at the micro-CT of each skull (Fig.~\ref{fig7_skull_microct}\flav{A}). \alex{While the skulls of sheep n$^{\textrm{o}}$4 and 5 show a regular thickness ($L\sim 4.2\pm 1.1$ mm and $L\sim5.3\pm 1$ mm, respectively)}, skull n$^{\textrm{o}}$6 displays a more irregular shape: $L \sim 5.7\pm 1.7$ mm~\cite{coudert_3d_2024}. This difference of morphology between the \alex{three} skulls \alex{could} explain the higher degree of aberrations observed \alex{in sheep n$^{\textrm{o}}$6}.

On the other hand, the multiple scattering rate can be estimated from the incoherent background of the RPSF (Methods, Eq.~\ref{multiple}). It appears that the multiple scattering rate is far from being negligible even after the beamforming process since it reaches a value of 30 \% in the $z=35$ mm-region for sheep n$^{\textrm{o}}$6 (Fig.~\ref{fig7_skull_microct}\flav{D}). As the aberration level, the multiple scattering rate is larger for this sheep over almost the whole depth range. This feature can be understood by looking at the internal structure of the bone. Indeed, the diplo\"{e} volume ratio $\nu$ is much higher in sheep n$^{\textrm{o}}$6 ($\nu=57$\%) than in \alex{sheeps n$^{\textrm{o}}$4 and 5 ($\nu=14$\% and $\nu=29$\%, respectively)}~\cite{coudert_3d_2024}. This high degree of heterogeneity seems to account for the higher multiple scattering rate observed in sheep n$^{\textrm{o}}$6. \\

\subsection*{Overcoming wave distortions}

\begin{figure}[tb!]
\includegraphics[width=\textwidth]{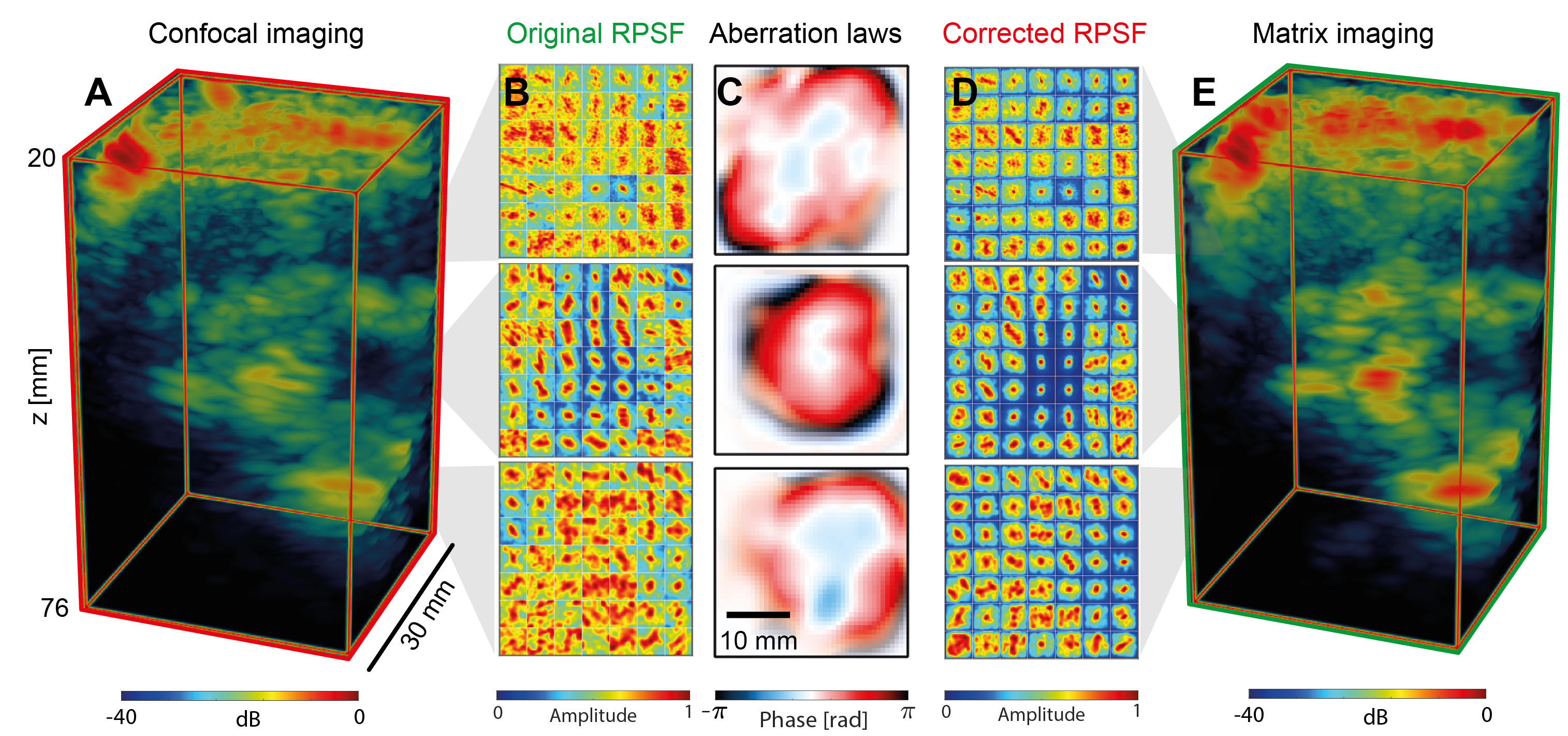}
    \caption{\textbf{Ultrasound matrix imaging in the sheep brain.} (\textbf{A}) Confocal volume extracted from the diagonal of the focused $\mathbf{R}-$matrices (Methods, Eq.~\ref{confocal}).   (\textbf{B}, \textbf{C}, and \textbf{D}) Maps of initial RSPFs, aberration phase laws, and corrected RPSFs, respectively,  at three different depths $z=32$ mm (top), $z=50$ mm (middle) and $z=68$ mm (bottom). Each RPSF is displayed in a scan range $\boldsymbol{\Delta \rho}=\boldsymbol{\rho}_\outm-\boldsymbol{\rho}_\inm$ (see Methods) that varies from $-15$ to $+15$ mm in both $x$ and $y$ directions. (\textbf{E}) Confocal image after aberration correction. The results shown here correspond to ultrasound data acquired on sheep n$^{\textrm{o}}$6.}
    \label{fig2_UMI_analysis}
\end{figure}

{The complex speed-of-sound distribution highlighted by Fig.~\ref{fig7_skull_microct}B hampers ultrasound brain imaging. The confocal image of the brain, that can be extracted from the diagonal coefficients of $\mathbf{R}_{\bm{\rho \rho}}(z)$ (Methods, Eq.~\ref{confocal}), is blurred by the skull heterogeneities (Fig.~\ref{fig2_UMI_analysis}A). Fortunately, an optimized contrast and diffraction-limited resolution can be recovered using the distortion matrix concept~\cite{lambert_distortion_2020,lambert_ultrasound_2022a,bureau_three-dimensional_2023}. This matrix contains the wave distortions exhibited by each reflected wave-front in the transducer basis (Fig.~\ref{fig1_UMI}C). The choice of this basis is dictated by the fact that most aberrations are induced by the skull heterogeneities, \textit{i.e} in the vicinity of the probe. An iterative phase reversal (IPR) process is then applied to the distortion matrix to provide an estimation of an aberration phase law, $\phi(\mathbf{u}_\outm,z)$, at each depth (Methods, Fig.~\ref{fig1_UMI}D and Movie S1). Examples of aberration phase laws are displayed in Fig.~\ref{fig2_UMI_analysis}C. They show a complex feature, as quantified by the Strehl ratio $\mathcal{S}$ that provides the ratio between the energy at focus with and without aberrations~\cite{mahajan1982} (Methods, Eq.~\ref{Strehl}). For instance, the aberration law measured at depth $z=50$ mm exhibits a Strehl ratio $\mathcal{S}=0.03$. This extremely low value illustrates the detrimental impact of skull heterogeneities on the focusing process. \alex{A more detailed quantification of the aberration magnitude is provided in Section~\ref{strehl_section} by evaluating the modulus of the aberration transfer function. As expected from the previous RPSF analysis (Fig.~\ref{fig7_skull_microct}), the aberration magnitude is found to be 30 \% larger in sheep n$^{\textrm{o}}$6 compared to sheep n$^{\textrm{o}}$4 and 5. }

The aberration phase laws can be leveraged to compensate for the wave distortions exhibited by the focused reflection matrix (Methods). The resulting confocal image is displayed in Fig.~\ref{fig2_UMI_analysis}E. It exhibits a clear improvement compared to the initial confocal image (Fig.~\ref{fig2_UMI_analysis}A), with a contrast gain of 8 dB for the deepest bright scatterers ($z=50$ and $60$ mm). The efficiency of the correction process can be assessed by comparing the maps of RPSFs before (Fig.~\ref{fig2_UMI_analysis}B) and after (Fig.~\ref{fig2_UMI_analysis}D) aberration compensation. The spatial resolution is improved by a factor between 2 and 3 and the multiple scattering background is decreased by more than 10 dB beyond $z=50$ mm (Fig.~\ref{figMS}). Nevertheless, the obtained RPSFs remain imperfect in Fig.~\ref{fig2_UMI_analysis}D. High-order aberrations subsist because of their anisoplanicity. Last but not least, even though the confocal image shows a clear improvement, the diffraction-limited resolution $\delta \rho_0$ remains limited by the probe size $D$: $\delta \rho_0 \sim \lambda z/D$. } \\

\subsection*{Matrix imaging for ultrasound localization microscopy}

\begin{figure}[ht]
\centering
\includegraphics[width=\textwidth]{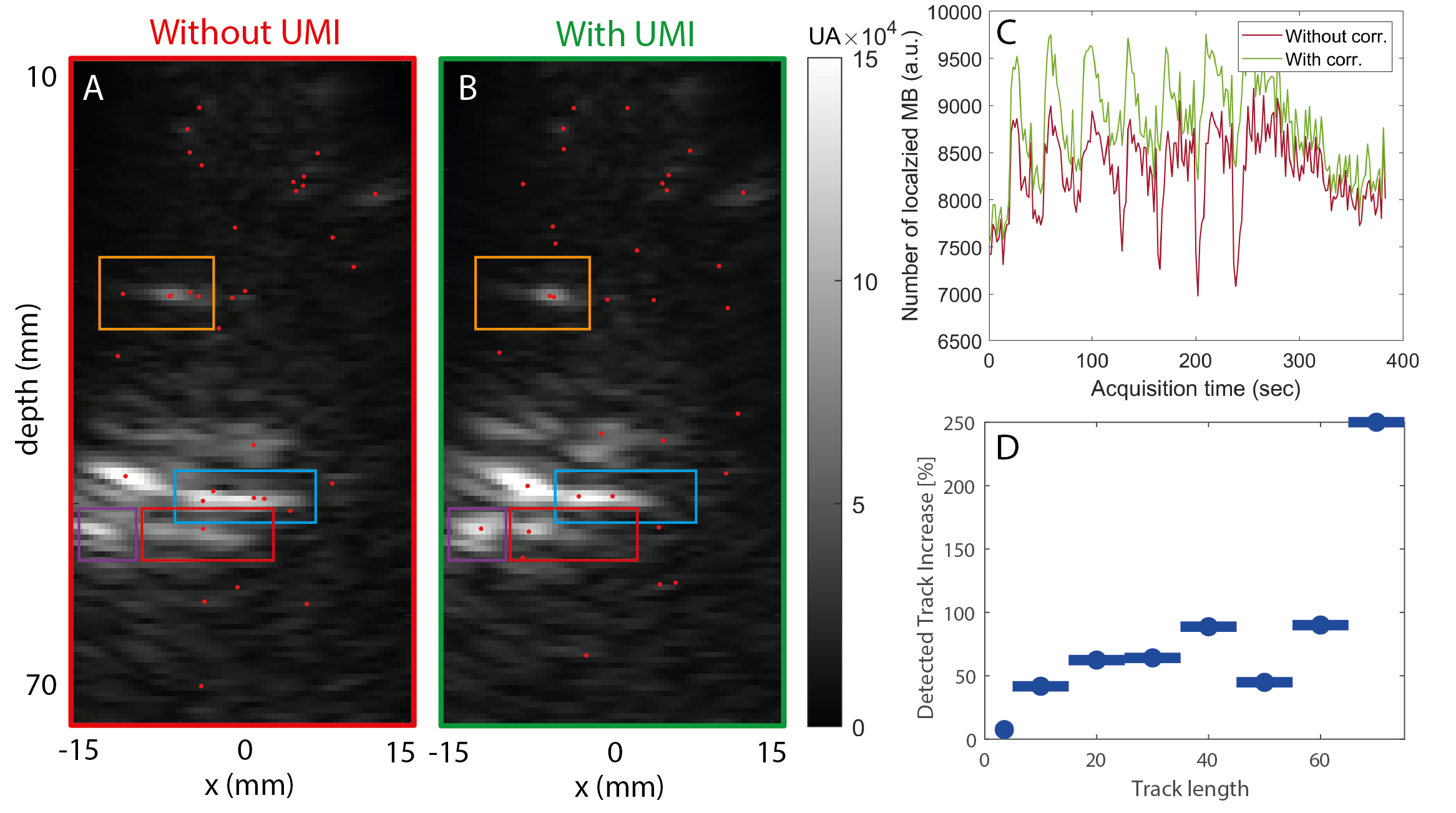}
\caption{\textbf{Enhancing micro-bubble localization accuracy with UMI.} (\textbf{A} and \textbf{B}) \flav{Sagittal} middle slice of the ultrasound image without and with aberration correction at a given time, respectively. The position of the detected microbubbles is highlighted by red dots. The blue and orange boxes show areas where false alarms are reduced by reducing the side lobes of the PSF. The purple region shows a region where aberration compensation allows the detection of a microbubble. The red square highlights a zone where a microbubble is reassigned at a new location. (\textbf{C}) Number of localized microbubbles as a function of the acquisition time $t$ with (green) and without (pink) UMI. (\textbf{D}) Increase of the microbubble detection rate with UMI for different track lengths. The data shown here correspond to sheep n$^{\textrm{o}}$ 6 (acquisition 4).}
\label{fig4_localization}
\end{figure}

{To overcome diffraction, ULM is then performed by injecting boluses of microbubbles and tracking them using an ultra-fast imaging sequence (Methods). To that aim, an hybrid transmit basis is used with three cylindrical waves emitted by the entire probe and two spherical diverging waves transmitted successively by each probe panel~\cite{coudert2024}. This sparse emission basis, referred to as $\mathbf{s}_\textrm{in}$, yields a volume rate of 209 Hz (Methods). For each incident wave-field, the raw data is stored in a set of new reflection matrices $\mathbf{R}^{'}_{\mathbf{us}}(\tau,t_m)=[R'(\mathbf{u}_\textrm{out},\mathbf{s}_\textrm{in},\tau,t_m)]$ recorded at different times $t_m$. A set of three-dimensional images is then obtained by beamforming the reflection matrix rephased by the conjugate of the aberration law to compensate for skull-induced aberrations:
\begin{align}
\mathcal{I}'(\mathbf{r},t_m)= \sum_{\mathbf{s}_\textrm{in}}  \sum_{\mathbf{u}_\textrm{out}} & \exp\left [ -j \phi(\mathbf{u}_\textrm{out},z) \right ] \nonumber \\
&\times  R'(\mathbf{u}_\textrm{out},\mathbf{s}_\textrm{in},\tau_{\textrm{in}}(\mathbf{s}_\textrm{in},\bm{\rho},z)+\tau_{\textrm{out}}(\mathbf{u}_\textrm{out},\bm{\rho},z),t_m)
\label{confocal_eq}
\end{align}
After filtering out the static component in the image series (Methods), the dynamic component of the brain can be highlighted. A two-dimensional cross-section of the resulting image is displayed in Fig.~\ref{fig4_localization}B at a given time $t_m$. This image is compared to its raw counterpart (Fig.~\ref{fig4_localization}A) obtained without aberration compensation. Thanks to UMI, isolated microbubbles show less distorted focal spots, as shown by two examples in the purple and red squares in Figs.~\ref{fig4_localization}A,B. Just above ($\sim$50 mm), a halo of microbubbles in the blue square seem to indicate the presence of a vessel but the individual detection of each microbubble is complicated by their high concentration.}

A deconvolution of the dynamic image with a theoretical PSF and the application of a detection threshold provides a localization of bubbles in the field-of-view  (Methods). The position of the detected bubbles are highlighted with red \flav{points} in Fig.~\ref{fig4_localization}A and B. While several microbubbles are wrongly assigned to the high side-lobes of each distorted focal spot in Fig.~\ref{fig4_localization}A, UMI drastically reduces this number of false detection in Fig.~\ref{fig4_localization}B by restoring a satisfying PSF, in particular in the colored squares. Despite this reduction of false alarms, the number of localized microbubbles is increased. Figure \ref{fig4_localization}C illustrates this assertion by showing, for sheep n$^\textrm{o}6$, the higher number of detected bubbles over time with UMI. The different peaks corresponds to each injection of microbubbles. 

To further reduce the false alarm rate and better appreciate the effect of UMI, microbubbles can be tracked over time (Methods). \alex{In absence of aberration correction, the focal spot is distorted and bubbles cannot be detected on each frame during their journey across the field-of-view. The bubble trajectory is then subdivided into shorter tracks. As a consequence, the track length is a relevant observable to demonstrate the benefit of the aberration correction process~\cite{robin_vivo_2023}.} Fig.~\ref{fig4_localization}D shows the increase of detected tracks provided by UMI as a function of their length. UMI significantly increases the number of detected tracks by a factor ranging from 40 to 250\% according to the track length. Table~\ref{Tab1LD} shows the same observable for different minimum path lengths. The general trend is the same for each acquisition, thereby showing the robustness of UMI and its benefit in each imaging configuration. \alex{This result is crucial for ULM images as the longest tracks are the most reliable to delineate small vessels.}
\begin{table}[ht]
\tiny
\resizebox{\textwidth}{!}{%
\begin{tabular}{|c|c|c|c|c|c|c|}
\hline
\multirow{2}{*}{\begin{tabular}[c]{@{}c@{}}Sheep \\ number\end{tabular}} & \multirow{2}{*}{\begin{tabular}[c]{@{}c@{}}Acq. \\ number\end{tabular}} & 
\multicolumn{5}{c|}{Additional tracks detected with UMI correction [\%]} \\
 &  &  $\geq$ \textbf{2} frames &  \textgreater \textbf{3} frames &  \textgreater \textbf{5} frames &  \textgreater \textbf{10} frames &  \textgreater \textbf{30} frames\\ \hline
S4 & 1 & 24.0 & 41.7 & 49.1 & 63.3  & 121.7 \\ \hline
\multirow{2}{*}{S5} & 1 & 14.1 & 26.6& 35.7 & 54.5  & 76.6 \\ 
                    & 2 & 7.5 & 13.7 & 16.2 & 28.5   & 42.8 \\ \hline
\multirow{5}{*}{S6} & 1 & 15.4 & 16.3 & 19.2 & 23.5   & 43.5 \\ 
                    & 2 & 8.7 & 13.3 & 15.3 & 19.1   & 36.1 \\ 
                    & 3 & 6.7 & 11.5 & 13.2 & 17.3   & 18.3 \\ 
                    & 4 & 6.9 & 22.3 & 32.5 & 47.2   & 66.2 \\ 
                    & 5 & 8.7 & 27.9 & 32.5 & 53.5   & 77.1 \\ \hline
\multicolumn{2}{|c|}{{Mean}} & {11.5} & {21.7} & {26.7} & {38.4} & {60.3} \\ 
\multicolumn{2}{|c|}{{Std}}  & {6.0} & {9.0}  & {11.2}  & {16.1} & {28.5} \\ \hline
\end{tabular}%
}
\caption{\textbf{Quantification of UMI benefits on ULM tracks.} Percentage of additional tracks provided by UMI for various minimal tracks length expressed in frames.}
\label{Tab1LD}
\end{table}

\begin{figure}[ht]
\includegraphics[width=\textwidth]{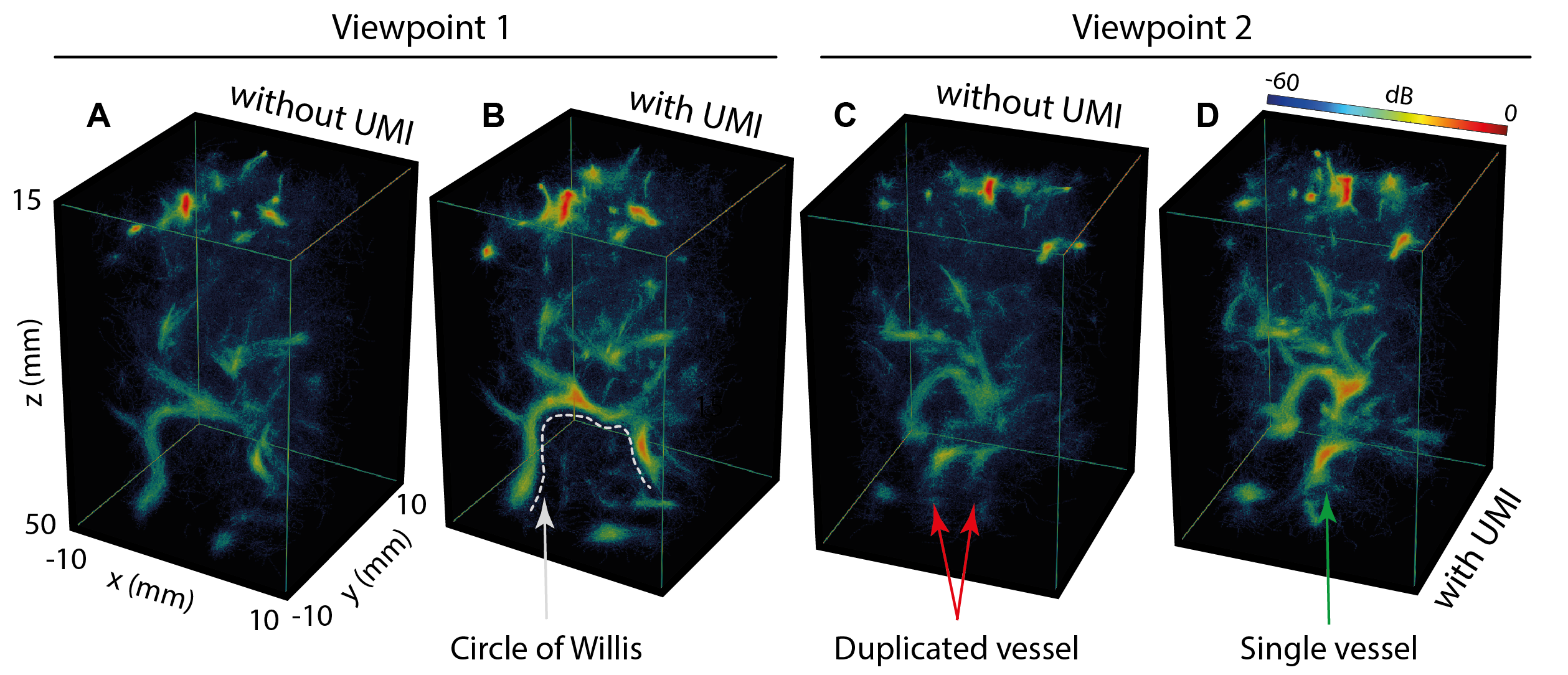}
    \caption{\textbf{Correcting ULM images with UMI.} 3D images shown as maximum intensity projections from two different points of view. ULM volumes (\textbf{A}, \textbf{C}) before and (\textbf{B}, \textbf{D}) after aberration compensation. \alex{The polygon of Willis is depicted by a dashed white line and} is located around $z=40-45$ mm. The same dynamic range is displayed for all images. The data shown here correspond to sheep n$^{\textrm{o}}$ 6, acquisition 4.}
    \label{fig3_ULMvolume}
\end{figure}

ULM density maps are built by counting the number of tracks crossing each voxel of the field of view. Such volumes are shown from two different angles in Fig.~\ref{fig3_ULMvolume}. The typical structure of the cerebral vessels can already be recognized before aberration correction (Figs.~\ref{fig3_ULMvolume}A and C). This structure is known as the Willis circle \cite{kalsoum_circle_2014} and can be seen as the main cross-road of the main vessels in the brain.

However, the skull heterogeneities also induce artifacts such as the duplication of one artery that can be noticed at the bottom of the conventional ULM image (Figs.~\ref{fig3_ULMvolume}A and C). This type of artifact has already been observed in transcranial imaging in both mouse~\cite{xing_phase_2023} and human~\cite{robin_vivo_2023}. Strikingly, the focusing laws extracted by UMI (Fig.~\ref{fig2_UMI_analysis}C) allow the removal of such artifacts as highlighted by the corrected ULM images in Figs.~\ref{fig3_ULMvolume}B and D. More generally, the higher detection rate and correct repositioning of micro-bubbles lead to a better contrast for ULM as highlighted by the vessel network on top of the Willis circle ($z=$25-35 mm) that appears much brighter (+6 dB) and better resolved. \alex{A similar} resolution and contrast improvement can be observed for the other 3D ULM images shown in Fig.~\ref{fig_supp1_ulm_volume} and corresponding to other acquisitions on the three sheeps.

\begin{figure}[ht]
\centering
\includegraphics[width=15cm]{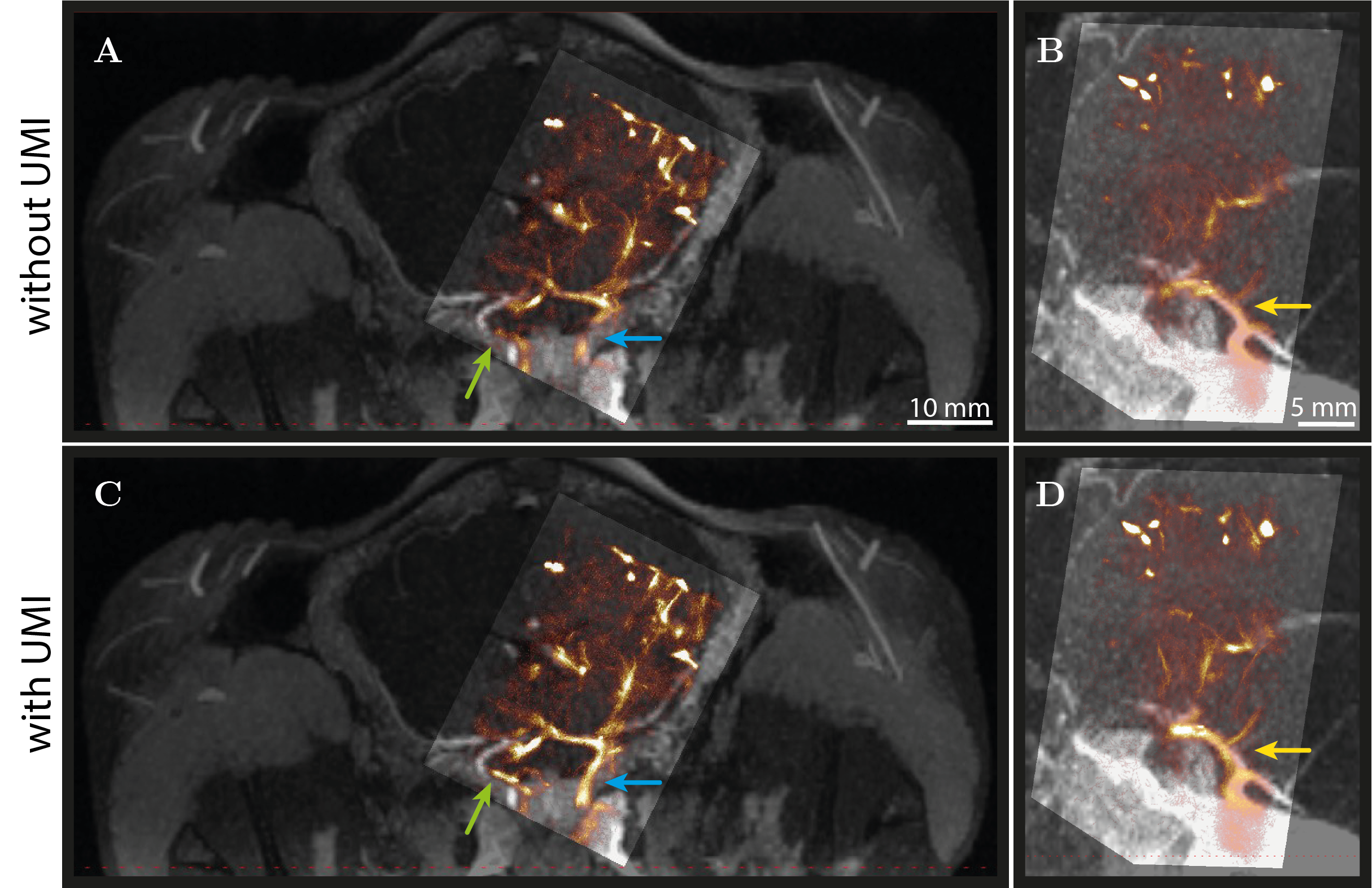}
\caption{ \textbf{Confronting ULM density maps to MRA images.} (\textbf{A},\textbf{B}) Coronal slices and sagittal slices, respectively: The ULM sections are shown as maximum intensity projections of 5 mm (coronal) and 1 cm (sagittal) thickness. Arbitrary colormaps are used with fixed dynamic range for the ULM (hot colormap) and MRA (B\&W scale) images to facilitate their comparison.  (\textbf{C},\textbf{D}) Same images as in panels A and B with ULM images corrected by UMI. Green and blue arrows show a better reconstruction of the posterior cerebral arteries in deviation from the Willis polygon. Yellow arrows show the correction in the venous system. The data shown here correspond to sheep number 6 (acquisition 4).}
\label{fig6_mra}
\end{figure}

To appreciate more quantitatively the benefit of UMI, the ULM density maps are superimposed on MRA, which is considered as the gold standard for observing large vascular structures in the brain. After UMI, the ULM density maps match more closely with the vessels revealed by MRA. In the coronal slice (Fig.~\ref{fig6_mra}A,C), UMI allows a more complete and brighter reconstruction of the vascular network (Fig.~\ref{fig6_mra}C) that matches in depth with the Willis polygon revealed by MRA, especially in the anterior cerebral arteries that deviate from the polygon (green and blue arrows). \alex{A zoom on the right anterior cerebral artery is displayed in Fig.~\ref{figComparison} and highlights the drastic gain in contrast provided by ULM compared to MRA around the Willis circle. This benefit of UMI can be quantified by a loss of image entropy~\cite{Tsai2007} increased by a factor 5 with UMI.}

In the sagittal section  (Fig.~\ref{fig6_mra}B,D), ULM provides a better adequation between MRA and ULM images of venous system (yellow arrows). \alex{This observation can be quantified by a computing structural similarity index between MRA and ULM images. This similarity metrics is increased by a factor 2 with UMI (Fig.~\ref{figComparison0}).} Besides the repositioning of large vessels, the increase in the total SNR of the ULM volume and the appearance of small vessels provided by UMI are noteworthy compared to the initial ULM image but also with respect to MRA that only provides an image of the largest vessels \alex{(Fig.~\ref{figComparison})}.

\alex{Finally, note that the volume rate (209 Hz) of our system limits the ability of the ULM pipeline to track microbubbles traveling in the larger cerebral vessels and to provide an appropriate flow velocity map~\cite{demene_transcranial_2021}. In the future, a 3D ultrasound scanner with a higher volume rate could alleviate this issue and provide additional physiological information.}

\section*{DISCUSSION}

Compared to a previous study on a static brain phantom~\cite{bureau_three-dimensional_2023}, we have here demonstrated the benefit of UMI for transcranial in-vivo imaging. Besides improving the contrast and resolution of standard ultrasound brain images, we also showed that it can be fruitfully coupled to other imaging modalities such as ULM. While the gain in contrast provided by UMI leads to increase the overall number of detected microbubbles, it also drastically increased the length of the microbubble tracks, which is affected by SNR, point-source separability and localization precision. Because UMI reduces the side lobes of the imaging PSF, the localization of microbubbles is actually much sharper. Doing so, UMI provides a more detailed ULM image and the removal of common artifacts such as duplication of vessels. The success of UMI has been confirmed by confronting ULM images with gold standard MRA.

More fundamentally, UMI also showed some correlation between the aberration level and the skull thickness variations. It also highlighted the impact of diplo\"{e} volume fraction on the multiple scattering level. The striking result of UMI on sheep n$^{\textrm{o}}$6 is promising for applications of UMI on human brain imaging, since this skull heterogeneity is in line with what is expected for human brain imaging across the temporal window~\cite{boruah_variation_2015, adanty_cortical_2021}. 

UMI approach for aberration compensation is perfectly complementary with the alternative route proposed recently by Robin \textit{et al.}~\cite{robin_vivo_2023} that exploited bubbles as ultrasound guide stars for trans-cranial adaptive focusing. To some extent, UMI is more robust since it directly synthesizes virtual guide stars from speckle~\cite{lambert_distortion_2020} and can work whatever the concentration of bubbles \alex{as well as in stronger aberration and multiple scattering conditions. Typically, we were not able to apply the approach proposed in Ref.~\cite{robin_vivo_2023} due to the low quality of the ULM data. Indeed, the latter method requires the presence of bubbles whose associated focal spot displays a correlation degree $\gamma$ larger than 0.6 with the ideal diffraction-limited PSF. Here, the normalized Strehl ratio, whose value is a maximal bound for $\gamma$, ranges from 0.1 to 0.4 (Section~\ref{strehl_section}). Such an aberration level makes difficult the use of microbubbles as guide stars in an adaptive focusing scheme. In Section\flav{~\ref{other}}, we also show how UMI provides a flexible benchmarking framework to show the benefits of the iterative phase reversal process~\cite{bureau_three-dimensional_2023} compared with conventional aberration correction methods in speckle such as iterative time reversal~\cite{montaldo_time_2011} or SVD beamforming~\cite{bendjador_svd_2020,Bendjador2021}.}

\alex{Nevertheless,} in the current realization, UMI also showed some limits. First, it requires the prior recording of a high-dimension reflection matrix. Second, in the present case, it only provided the depth-dependence of aberration phase laws and was not able to capture the lateral variations of aberrations. Last but not least, it did not compensate for the multiple reverberations inside the skull that prevented us from brain imaging right below the skull. \alex{Those limits are explained by the fact that high-order aberrations and reverberations are associated with extremely small isoplanatic patches. This feature limits the efficiency of the spatial averaging process required for the synthesis of a coherent guide star in matrix imaging (Fig.~\ref{fig1_UMI}).} 

In future works, these limits will be addressed by exploiting the dynamics of blood vessels. Indeed, rapid decorrelation of speckle may offer an opportunity as it provides numerous speckle realizations for a single voxel. Therefore, the aberration phase laws could in principle be extracted at a higher spatial resolution, since UMI would no longer have to rely on the isoplanicity assumption~\cite{zhao_phase_1992,osmanski_aberration_2012}. In other words, time averaging will replace spatial averaging, and the aberration phase laws can be extracted directly from the ultrafast sequence [Fig. \ref{fig1_acquisition}D] and not from a prior static sequence [Fig. \ref{fig1_acquisition}C]. This access to a large number of disorder realizations can also be leveraged for tailoring complex spatio-temporal focusing laws capable of harnessing multiple reverberations.  

In the future, we therefore expect that aberration and scattering compensation provided by UMI could help ULM in the detection and classification of cerebrovascular accidents in humans. An important goal will be to distinguish between ischemic and hemorrhagic strokes in the early phase, as has been done in the rat brain~\cite{chavignon_3d_2022}. Therefore, one perspective of this work is to conduct a pilot study with patients who have recently suffered a stroke.

In this paper, we have shown the great advantages and relative simplicity of the UMI and ULM combination. 3D transcranial in-vivo imaging of sheep brain has been performed and shown to be in excellent agreement with gold standard MRA. Although we have proven this claim for the specific case of ultrasound data acquired with a multiplexed matrix probe, these results are much more general. Indeed, UMI can be applied to any other ultrasound probe (e.g. sparse or RCA array), any other modality (e.g. conventional Doppler or shear wave elastography) and in either 2D or 3D configurations.

Beyond the specific case of ultrasound, this study paves the way towards the combination of matrix imaging~\cite{Balondrade2024,Giraudat2024} with super-resolution localization techniques~\cite{betzig_imaging_2006} in all fields of wave physics, ranging from optical microscopy with fluorescent molecules~\cite{Park2023} to seismology for dynamic imaging of glaciers~\cite{Nanni2021,Nanni2022}.

\section*{MATERIALS AND METHODS}
\subsection*{Experimental procedure}
All experiments were performed between October 2022 and January 2023 in accordance with ARRIVE guidelines, the European Directives and the French Legislation on Animal Experimentation (\#32738) and approved by the local ethics committee (CENOMEXA, No54). A total of 6 female domestic sheep were included in the study (Ovis aries weighing 35-40 kg and aged 15-16 months). Three of them were used to optimize the experimental protocol, the other three are presented here (Table~\ref{tabAcq}).

The sheep were sedated with ketamine (12 mg/kg) and xylazine (1 mg/kg) for placement of the venous catheter. Propofol (100 $\mu$g/kg) was then administered and endotracheal intubation was performed. During the surgical and imaging procedures, anesthesia was maintained by administration of sevoflurane (1.5-2.0 \%) in air. Cardiac activity, oximetry and arterial pressure (between 80 and 120 mm Hg) were monitored throughout the procedure. Respiratory rate and tidal volume were adjusted to maintain physiologic limits, and body temperature was maintained at 38 $^{\textrm{o}}$C with a warming blanket.

At the end of the ultrasound examination, i.e. after 5 hours, the anesthesia was terminated so that the animal could wake up and be returned to its box. Three imaging sessions were performed within one to two weeks. At the end of the last imaging session, the sheep was euthanized with a barbiturate overdose (pentobarbital 0.2g/kg) after 2 minutes of 8\% sevoflurane inhalation. A controlled arterial pressure of less than 20 mmHg and no cardiac activity were used to assess death.

After the last imaging session and the death of the animal, the part of the skull below the probe was removed for cleaning and preserved in saline and bleach. This subject was then fixed in a bath of formalin-free fixative F13 (Morphisto, Germany) for one day, then dried and placed under vacuum for micro-CT imaging (Skyscan 1176, Bruker, Kontich, Belgium, voxel size 18 $\mu$m)~\cite{coudert_3d_2024}.\\

\subsection*{Ultrasound sequences}
A low-frequency matrix array probe (Vermon, Tours, France), as described in Table~\ref{probecouture}, is placed on the shaved head of the sheep directly over the crest of the frontal bone. 3D ultrasound sequences were then recorded with a Vantage 256 4-to-1 multiplex system (Verasonics UTA 1024-MUX adapter, Kirkland, USA). The 1024 elements were connected to the 256 channels of the ultrasound scanner using a 4-to-1 multiplexer (Verasonics UTA 1024-MUX Adapter). \alex{The probe is therefore divided into four panels, the multiplexer only allowing emission with one panel and receiving with another~\cite{coudert_3d_2024,coudert2024}. For the reflection matrix acquisition, a set of 16 emissions had to be performed for any incident wave-field in order to record the $4 \times 4$ responses between each panel at emission and reception. Those recordings are then reassembled in post-processing to get the full reflected-wave-front associated with each synthetic incident wave-field. For the ULM sequence, this constraint has been partially mitigated since it is possible to emit with the four connected panels at the same time, provided that the emission waveform and delays were identical along the elevation axis~\cite{coudert_3d_2024,coudert2024}. As detailed further, the use of incident cylindrical wave-fronts was therefore used to reach a sufficient frame rate for ULM.} The successive ultrasound imaging sequences of the entire sheep experiment are summarized in Fig.~\ref{fig1_acquisition}.
\begin{table}[!ht]
    \center
\begin{tabular}{|l||l|}
    \hline
  {Number of transducers} & $32 \times 32=1024$ (with 5 dead elements)\\
  \hline
  {Geometry} (y-axis) & 3 inactive rows between each block of 32x8 elements\\
      \hline
   {Pitch} & $\delta u=0.55$ mm ($\approx \lambda/2$ at $c=1540$ m/s)\\
   \hline
   {Aperture} & $\Delta u =\begin{pmatrix}
        \Delta u_x \\  \Delta u_y
    \end{pmatrix} = \begin{pmatrix}
        17.6\textrm{ mm} \\  19.3\textrm{ mm}
    \end{pmatrix} $\\
  \hline
  {Central frequency} & $f_\textrm{c}=1.56$ MHz\\
  \hline
   {Bandwidth} & $60\% \rightarrow [1-2]$ MHz\\
    \hline
   {Transducer directivity} & $\theta_{max} =  64^{\circ}$ at  $c=1540$ m/s\\
   \hline
\end{tabular}
\caption{\textbf{Matrix array datasheet.}}
\label{probecouture}
\end{table}

Initial real-time power Doppler imaging is used to correctly position the probe before acquiring longer ultrasound data. This sequence is performed with the injection of microbubble contrast agent (Sonovue, Bracco, Italy) \cite{schneider_characteristics_1999} to obtain SNR-enhanced images of the cerebral vasculature. Once a sufficiently good signal is obtained, the probe is fixed, and we wait until the microbubbles has disappeared, i.e. 10 minutes.

A UMI sequence is then performed. It consists in the acquisition of the reflection matrix using a set of 324 synthetic spherical waves whose associated virtual sources are located along a sphere radius $\mathcal{R}=39.5$ mm: $\mathbf{R}_{\mathbf{uv}}(t)=[R(\uout,\mathbf{v}_{\inm},t)]$, where the vector $\mathbf{v}_{\inm}$ indicates the position of the corresponding virtual sources (see Supplementary~\ref{fig_supp6_virtual_sources}) and $\uout$ refers to the position of the transducers. The insonification of the medium is made with diverging waves associated with virtual sources located above the probe~\cite{provost_3d_2014} such that: 
\begin{equation}
    \vg=\begin{pmatrix}
        v_x \\ v_y \\  v_z
    \end{pmatrix}  = -\mathcal{R}\begin{pmatrix}
        \sin \theta_x \\ \sin \theta_x \\  \sqrt{1-  \sin^2 \theta_x- \sin^2 \theta_y}
    \end{pmatrix},
    \label{virtualsourceposition}
\end{equation}
with $\bm{\theta}=(\theta_x,\theta_y)$ the  angle associated with each virtual source. \alex{The angular range spans from $-\sin \theta_m$ to $+\sin \theta_m$, with $\theta_m=30^\textrm{o}$. The angular pitch $\delta (\sin \theta)$ is governed by the probe extension: $\delta (\sin \theta)=\lambda_c /\Delta u$ , with $\lambda_c=c_0/f_c$, the central wave length. This angular range and angular pitch dictate the number of incident diverging waves: $N=(2 \sin \theta_m /\delta (\sin \theta) )^2\sim 324$.} 

Finally, a hybrid ultrasound localization microscopy (ULM) sequence is performed. \alex{To limit the number of incident waves, only the azimuth and elevation shifts $(\theta_x,\theta_y)$ of {(0,0); (-5, 0); (5, 0); (0, -5); (0, 5)} degrees are considered. As stated above, the 4-to-1 multiplexer imposes a sequence of 16 insonifications to record the response of the medium for each virtual source. Such a sequence is prohibitive for high-frame rate imaging. However, this condition can be relaxed by only recording the responses between neighboring panels, thereby restricting the number of insonifications to 10 for each virtual source~\cite{coudert2024}. Moreover, the four panels can be connected together for the emission if the delays are similar in the elevation direction. For such an incident cylindrical wave-front, only 4 successive transmissions are required for the 4 reception apertures. However, full cylindrical waves could only be induced when the angle $\theta_x$ is 0 in the azimuthal direction. In the $y-$direction, waves still needed to be split in multiple emission-reception. Consequently, an hybrid sequence was chosen to optimize the frame rate while maintaining a sufficient angular diversity for the acquisition: 
\begin{itemize}
\item Three cylindrical waves were emitted sequentially from the entire probe and had to be repeated four times to record the reflected wave-front on each panel. 
\item Two spherical waves (virtual source) were decomposed by panels, corresponding to 10 emissions. 
\end{itemize}
The full ULM sequence thus corresponds to 32 emissions to achieve a volume rate of 209 Hz\cite{coudert2024, coudert_3d_2024}}. Since only a few insonifications are used, the ULM acquisition consists of a partial reflection matrix that is acquired at \alex{multiple times} $t_{m}$ so that the reflected wavefronts are stored in a second matrix $\mathbf{R}'=[R'(\uout,\mathbf{i}_\inm,\tau,t_{m})]$. \alex{The sparse feature of the recorded reflection matrices implies an ultrasound image of lower quality compared to a full measurement of the $\mathbf{R}$-matrix. Both the contrast and, to a lessen extent, the resolution of the beamformed image are degraded. However, those detrimental effects are compensated, at least partially, by the echogenicity of the bubbles for the contrast and by the localization feature that allows the resolution to be no longer limited by diffraction but by the signal-to-noise ratio~\cite{desailly_resolution_2015}.} During this acquisition, microbubble injection boluses (Sonovue, Bracco, Italy) were injected, i.e. from $1.5$ to $2$ ml/min, during 2.5 to 6.5 minutes. 

A total of 1 to 6 acquisitions were performed per sheep. The experimental conditions are summarized in Table~\ref{tabAcq}. \alex{The raw ULM data used in the paper corresponds to the ULM data presented in Ref.~\cite{coudert_3d_2024}.} It should \alex{also} be noted that, for some acquisition, the UMI sequence is performed after the ULM sequence and not before, but this does not change the principle of the method.

\subsection*{Focused reflection matrix}

A general procedure to build each focused reflection matrix, noted $\mathbf{R}_{\bm{\rho\rho}}(z)$, is to use a delay-and-sum (DAS) algorithm, which consists of applying appropriate time delays to create a double synthetic focusing procedure at both input and output (Bureau 2023)\cite{bureau_three-dimensional_2023}. Mathematically, such a beamforming procedure is described by Eq.~\ref{generalbemaformchap5}.
The input and output times-of-flight, $\tau_{\textrm{in}}$ and $\tau_{\textrm{out}}$, are given by: 
\begin{equation}
\label{travel_time}
    \tau_{\textrm{in/out}}(\mathbf{w},\mathbf{r})=\frac{|\mathbf{w}-\mathbf{r}|}{c_0}=\frac{\sqrt{(x-w_x)^2+(y-w_y)^2+(z-w_z)^2}}{c_0}.
\end{equation}
with $\mathbf{w}=\mathbf{v}_\inm$ at input and $\mathbf{w}=\mathbf{u}_\outm$ at the output.

The confocal volume shown in Fig. \ref{fig2_UMI_analysis}e can be extracted from the diagonal elements of the reflection matrix ($\rhoin=\rhoout$), so that :
\begin{equation}
\label{confocal}
    \mathcal{I}(\bm{\rho},z)=\left\lvert R(\bm{\rho},\bm{\rho},z)\right\rvert^2
\end{equation}

\subsection*{Reflection point spread function}

The off-diagonal points in $\mathbf{R}_{\bm{\rho}\bm{\rho}}(z)$ can be exploited for a quantification of the focusing quality at any pixel of the ultrasound image. To that aim, the spreading of energy is investigated along each antdiagonal of $\mathbf{R}_{\bm{\rho}\bm{\rho}}(z)$. Mathematically, this can be done by means of a change of variable in order to express the focused $\mathbf{R}$-matrix in a common mid-point basis~\cite{bureau_three-dimensional_2023}:
\begin{equation}
    R(\Delta \boldsymbol{\rho},\boldsymbol{\rho}_\textrm{m},z)= R(\bm{\rho}_\textrm{out},\bm{\rho}_\textrm{in},z)
\end{equation}
with $\bm{\rho}_\textrm{m}= (\bm{\rho}_\textrm{in}+\bm{\rho}_\textrm{out})/2$, the mid-point between $\bm{\rho}_\textrm{in}$ and $\bm{\rho}_\textrm{out}$, and $\Delta \boldsymbol{\rho}=\bm{\rho}_\textrm{out}-\bm{\rho}_\textrm{in}$, their relative position. A local average of the back-scattered intensity is then performed needed to smooth out the speckle reflectivity and provide an estimator of the reflection point spread function (RPSF):
\begin{equation}
\label{RPSF}
    RPSF(\Delta \boldsymbol{\rho},\mathbf{r}_\textrm{p})=\left \langle \left\lvert R(\Delta \boldsymbol{\rho},\bm{\rho}_\textrm{m},z)\right\rvert^2 \mathcal{P}(\bm{\rho}_\textrm{m}- \bm{\rho}_\textrm{p},z-z_\textrm{p}) \right \rangle_{ \rho_\textrm{m},z }
\end{equation}
with $\mathbf{r}_\textrm{p}=(\bm{\rho}_\textrm{p},z_\textrm{p})$. The symbol $\langle \cdots \rangle$ denotes a spatial average {over the variables in the subscript}. $\mathcal{P}(\bm{\rho}_\textrm{m}- \bm{\rho}_\textrm{p},z-z_p) = 1$ for $|\boldsymbol{\rho}_\textrm{m} - \rhog_{\textrm{p}}|<p_\rho {/2}$ and $|z_\textrm{m}-z_{\textrm{p}}|<p_z{/2}$, and zero otherwise. The dimensions of $\mathcal{P}$ used for Figs.~\ref{fig2_UMI_analysis}c,d and \ref{fig7_skull_microct}e-h are $\mathbf{p}=(p_\rho,p_z)=(4.3,3)$ mm.\

As shown by Figs.~\ref{fig2_UMI_analysis}c,d and Figs.~\ref{fig7_skull_microct}e-h, the RPSF displays the following typical shape: A distorted confocal spot associated with singly-scatterered echoes on top of a flat background induced by multiple scattering events. On the one hand, the extension $\delta \rho $ of the confocal peak is directly related to the local resolution of the ultrasound image. $\delta \rho $ can be obtained by considering the de-scanned area $\mathcal{A}_{-3 dB}$ over which the normalized RPSF is larger than -3dB (Supplementary Fig.~\ref{fig_supp3_resolution}), such that 
\begin{equation}
\label{resolution}
\delta \rho  =\sqrt{\mathcal{A}_{-3 dB}/\pi}.
\end{equation}
On the other hand, the multiple scattering rate $\alpha_M$ can be estimated by averaging the RPSF beyond the confocal peak, \textit{i.e} for $4 \delta \rho_0<||\Delta \bm{\rho}|| <6 \delta \rho_0$:
\begin{equation}
\label{multiple}
\alpha_M (\mathbf{r}_{\textrm{p}}) =\frac{\left \langle RPSF (\Delta \bm{\rho},\mathbf{r}_{\textrm{p}}) \right \rangle_{4 \delta \rho_0<||\Delta \bm{\rho}|| <6 \delta \rho_0} }{RPSF (\Delta \bm{\rho}=\mathbf{0},\mathbf{r}_{\textrm{p}}) }
\end{equation}\\

\subsection*{Aberration phase law}

An aberrations law is extracted using the distortion matrix framework~\cite{lambert_distortion_2020,bureau_three-dimensional_2023} following these different steps:  
\begin{itemize}
    \item[--] (i) \alex{We first project} $\R_{\bm{\rho}\bm{\rho}}(z)$ onto the transducer basis ($\mathbf{u}$) either at input or output using a homogeneous propagation model:
    \begin{equation}
 \mathbf{R}_{\mathbf{u}\bm{\rho}}(z)=\mathbf{G}_0(z) \times \mathbf{R}_{\bm{\rho}\bm{\rho}}(z),
 \end{equation}
 and
 \begin{equation}
 \mathbf{R}_{\bm{\rho}\mathbf{u}}(z)= \mathbf{R}_{\bm{\rho}\bm{\rho}}(z)  \times
 \mathbf{G}_0(z)^{\top}   
 \end{equation}
 with $ \mathbf{G}_0=[G_0(\mathbf{u},\bm{\rho},z)]$, the transmission matrix between the probe and the set of voxels at depth $z$ for an ideal homogeneous medium:
 \begin{equation}
T_0(\mathbf{u},\bm{\rho},z)=\frac{z \exp\left ( j k_0 \sqrt{||\mathbf{u}-\bm{\rho}||^2+z^2} \right )}{4\pi (||\mathbf{u}-\bm{\rho}||^2+z^2)}
 \end{equation}
 \alex{As sketched in Fig.~\ref{fig1_UMI}B, this dual reflection matrix contains the wave-fronts associated with virtual sources corresponding to the focusing points $(\bm{\rho}_{\textrm{in/out}},z)$. Those wave-fronts correspond to the sum of a geometric parabolic component related to to the position of each focusing point $(\bm{\rho}_{\textrm{in/out}},z)$ and of a distorted component resulting from the mismatch between the propagation model and the real speed-of-sound distribution. Interestingly, this distorted component displays strong correlations because each reflected wave-front goes through the same part of the skull, the probe being placed at the skull surface. This is the so-called angular (or tilt-tilt) memory effect~\cite{osnabrugge_generalized_2017} that we will exploit in the following.}
    \item[--] (ii) To do so, we isolate the distorted part of each wavefront by subtracting the geometric component that would be ideally obtained in absence of aberrations (i.e. constant speed of sound $c_0$). The result is the input and output distortion matrices which are expressed as follows:
 \begin{equation}
 \label{D1}
\mathbf{D}_{\mathbf{u}\bm{\rho}}(z)= \mathbf{R}_{\mathbf{u}\bm{\rho}}(z) \circ  \bar{\mathbf{T}}_0 
 \end{equation}   
 and, 
  \begin{equation}
   \label{D2}
\mathbf{D}_{\bm{\rho}\mathbf{u}}(z)= \mathbf{R}_{\bm{\rho}\mathbf{u}}(z) \circ \bar{\mathbf{T}}_0^{\top} 
 \end{equation} 
\alex{where the symbol $\circ$ stands for the Hadamard (element wise) product} and with, $\bar{\mathbf{T}}_0$, the normalized transmission matrix such that $$\bar{{T}}_0(\mathbf{u},\bm{\rho},z)={{T}}_0(\mathbf{u},\bm{\rho},z)/|{{T}}_0(\mathbf{u},\bm{\rho},z)|.$$
\alex{ In the focused basis, the operation described in Eqs.~\ref{D1} and~\ref{D2}  amounts to a de-scan of each virtual source $(\bm{\rho}_{\textrm{in/out}},z)$ at the origin $(\bm{0},z)$ of each focal plane, as sketched in Fig.~\ref{fig1_UMI}C. In the transducer basis, it consists in a realignment of associated wave-fronts, thereby highlighting the angular correlations between wave distortions. }  
    \item[--] (iii) \alex{We then exploit} this angular memory effect by computing the correlation matrices of wave-front distortions, 
     \begin{equation}
\mathbf{C}_{\textrm{in}}(z)= \mathbf{D}_{\mathbf{u}\bm{\rho}}(z) \times  \mathbf{D}_{\mathbf{u}\bm{\rho}}^{\dag}(z),
 \end{equation}   
and
  \begin{equation}
\mathbf{C}_{\textrm{out}}(z)= \mathbf{D}_{\bm{\rho}\mathbf{u}}^{T}(z)\times \mathbf{D}_{\bm{\rho}\mathbf{u}}^*(z).
 \end{equation}
   \item[--] (iv) A numerical iterative phase reversal \alex{(IPR)} algorithm~\cite{bureau_three-dimensional_2023} is performed to extract an estimation of aberration laws $\bm{\Phi}_{\inm/\outm}(z)=[{\phi}_{\inm/\outm}(\ugg_{\inm/\outm},z)]$ by solving the following equation:
   \begin{equation}
   \label{recursive}
   \bm{\Phi}_{\inm/\outm}(z)=\mbox{arg} \left  \lbrace \mathbf{C}_{\textrm{in/out}} \times \exp \left ( i \bm{\Phi}_{\textrm{in/out} }(z) \right) \right \rbrace .
   \end{equation}
\alex{The IPR process, sketched in Fig.~\ref{fig1_UMI}D, allows the synthesis of a coherent guide star from  the set of virtual sources $(\bm{\rho}_{\textrm{in/out}},z)$. The associated wave-front is the resulting aberration phase law $\bm{\Phi}_{\textrm{out/in} }(z)$.} The complexity of each aberration law can be quantified by the Strehl ratio which is defined as follows~\cite{mahajan1982} \alex{(see also Section~\ref{strehl_section})}:
\begin{equation}
\label{Strehl}
\mathcal{S}(z)=\left |\langle \exp \left [i\phi(\mathbf{u},z)\right ] \rangle_{\mathbf{u}}
\right |^2.
\end{equation} 
The aberration phase laws yield an estimator of the input/output transmission matrix between the probe and each focal plane:
\begin{equation}
\label{T}
\mathbf{T}_{\textrm{in/out}}(z)=\exp \left ( i \bm{\Phi}_{\textrm{out}}\right) \circ \mathbf{T}_0 
\end{equation}
    \item[--] (vi) Aberrations are compensated by applying the phase conjugate of the transmission matrices to provide an updated focused reflection matrix:
        \begin{equation}
        \label{cor1}
 \mathbf{R}_{\bm{\rho}\bm{\rho}}(z)=\mathbf{T}_{\textrm{in}}^\dag(z) \times \mathbf{R}_{\mathbf{u}\bm{\rho}}(z),
 \end{equation}
 and
   \begin{equation}
    \label{cor2}
 \mathbf{R}_{\bm{\rho}\bm{\rho}}(z)= \mathbf{R}_{\bm{\rho}\mathbf{u}}(z) \times \mathbf{T}_{\textrm{out}}^*(z),
 \end{equation}
\end{itemize}
The whole process is repeated two times at input and output to improve gradually the focusing process. \alex{The whole UMI process is summarized by the flow chart displayed in Fig.~\ref{fig_flowchart}.} The confocal volume (Eq.~\ref{confocal}) displayed in Fig.~\ref{fig2_UMI_analysis}f and the map of RPSFs (Eq.~\ref{RPSF}) shown in Fig.~\ref{fig2_UMI_analysis}d are extracted from the diagonal coefficients of this corrected focused reflection matrix.

\subsection*{Ultrasound localization microscopy}

From the reflection matrix $\mathbf{R'}$ recorded using the hybrid sequence, a ULM image can be built by performing the following post-processing steps. 
\begin{itemize}
    \item[--] (i) High volume rate ultrasound images are constructed using a conventional delay-and-sum algorithm into a 300$\times$300$\times$600 mm$^3$ volume with 0.5$\times$0.5$\times$0.5 mm$^3$ voxels~\cite{coudert2024}: 
\begin{equation*}
\mathcal{I}'(\bm{\rho},z,t_m)= \sum_{\mathbf{s}_\textrm{in}}  \sum_{\mathbf{u}_\textrm{out}}  R'(\mathbf{u}_\textrm{out},\mathbf{s}_\textrm{in},\tau_{\textrm{in}}(\mathbf{s}_\textrm{in},\bm{\rho},z)+\tau_{\textrm{out}}(\mathbf{u}_\textrm{out},\bm{\rho},z),t_m)
\end{equation*}
    \item[--] (ii) A high-pass time filter is applied to the set of images by means of a two frames sliding average window: \alex{
    \begin{equation*}
\mathcal{I}''(\bm{\rho},z,t_m)= \mathcal{I}'(\bm{\rho},z,t_m) - \left \langle \mathcal{I}'(\bm{\rho},z,t_{n}) \right \rangle_{n \in [m-1;m+1]}
\end{equation*}
    Albeit crude, this simple operation allows the removal of most of the slowly-varying component of the ultrasound image (\textit{i.e} tissues) while enhancing the dynamic part of the image (moving microbubbles).}
    \item[--] (iii) The microbubbles were then detected in each volume by localizing regional maxima on each image $\mathcal{I}\alex{''}$. The center of each PSF was sub-localized using the radial symmetry method \cite{heiles_volumetric_nodate,Heiles2022,parthasarathy_rapid_2012}, i.e. by assuming a PSF Gaussian shape and by applying a gradient algorithm. Microbubbles with a signal-to-noise ratio (SNR) below 10.5 \purple{dB} were discarded. This SNR was calculated as the intensity at the center of the microbubbles divided by the mean intensity of the neighboring voxels over a 5-mm-wide patch. The detected bubbles and their localization are superimposed to one cross-section section of the ultrasound image in Figs.~\ref{fig4_localization}a,b.
    \item[--] (iv) Tracking algorithm has been performed using the Hungarian method~\cite{tinevez_simpletracker_2019} to build the histogram displayed in Fig.~\ref{fig4_localization}d. A maximum linking distance between two microbubbles is fixed to 1.5 mm. A one frame gap is allowed and no restriction on track duration is imposed. 
    \item[--] (v) A 3D ULM density map $\mathcal{U}(\rg)$ is obtained by accumulating and projecting all tracks of microbubbles onto a three-dimensional grid whose spatial sampling is $\lambda$/10. The result is displayed in Fig.~\ref{fig3_ULMvolume}a,c.
\end{itemize}
The same process is applied to the ultrasound images $\mathcal{I}\alex{''}$ corrected by UMI (Eq.~\ref{confocal_eq}). The resulting ULM maps are displayed in Figs.~\ref{fig3_ULMvolume}b,d.\\

\subsection*{Computational Insights}

\alex{The computation time of the UMI process~\cite{bureau_three-dimensional_2023} is limited by the beamforming of the focused R-matrix (Eq.~\ref{generalbemaformchap5}) that takes here 30 min (on GPU with CUDA language) while the aberration correction process only lasts for a few minutes. All the post-processing was realized with Matlab (R2024a) on a working station with 2 processors @2.20GHz, 128Go of RAM, and a GPU with 48 Go of dedicated memory. This computation time for UMI remains acceptable compared to the ULM process that takes 1h30min for 9000 frames. \flav{It should be noted that the UMI process (beamforming and correction) can be further speed up by reducing the number of focusing points required to extract the aberration laws.}}

\subsection*{Angiographic MRI sequences} 

Angiographic MRI sequences (MRA) were performed with a 3T MRI scanner (GE SIGNA PREMIER) in a thermoregulated room, i.e. at approximately 20$^{\textrm{o}}$C (axial propeller, TR=5539 ms, TE=165.744 ms, slice thickness 2 mm, 31 slices, in-place resolution 0.375$\times$0.375). The superimposition of MRA and ULM images in Fig.~\ref{fig6_mra} has been performed manually using Amira software (Thermofischer, v2019.4). \\

\section*{SUPPLEMENTARY MATERIALS}

Section~\ref{virtual}. Virtual source basis.

\noindent \alex{Section~\ref{flowchart}. Flow chart of the UMI process.}

\noindent \alex{Section~\ref{strehl_section}. {Magnitude of aberration correction}.}

\noindent Section~\ref{resenh} Resolution enhancement.

\noindent Section~\ref{msr} Multiple scattering rate.

\noindent \alex{Section~\ref{other} {Comparison \flav{with} other aberration correction methods}.}

\noindent \alex{Section~\ref{comparison} {Quantitative comparison between MRA and ULM images}.}

\noindent Section~\ref{ulmall} ULM images for each dataset.

\noindent Fig.~\ref{fig_supp6_virtual_sources}. Virtual source basis.

\noindent \alex{Fig.~\ref{fig_flowchart}. Flow chart of the UMI process.}

\noindent \alex{Fig.~\ref{fig_supp_strehl}.{Quantifying the magnitude of aberration correction.}}

\noindent Fig.~\ref{fig_supp3_resolution}. Quantifying the ultrasound image resolution.

\noindent Fig.~\ref{figMS}. Single and multiple scattering rates.

\noindent \alex{Fig.~\ref{figCorrComparison}. Comparison with other aberration correction methods.}

\noindent \alex{Fig.~\ref{figComparison0}. Quantitative comparison between ULM and MRA cross-sections in  the saggital plane.}

\noindent \alex{Fig.~\ref{figComparison}. Quantitative comparison between ULM and MRA cross-sections in the coronal plane.}

\noindent Fig.~\ref{fig_supp1_ulm_volume}. Impact of UMI on different ULM  acquisitions.

\noindent Tab.~\ref{tabAcq}. Experimental conditions for each acquisition.

\noindent \href{https://www.science.org/doi/full/10.1126/sciadv.adt9778#supplementary-materials}{Movie S1}. Result of UMI at each depth

\noindent \href{https://www.science.org/doi/full/10.1126/sciadv.adt9778#supplementary-materials}{Movie S2}. ULM images for different acquisitions before and after UMI correction.

\noindent References~\cite{tanter_time_2000,Denis2025}.

\section*{REFERENCES AND NOTES}

\bibliographystyle{ScienceAdvances}

\noindent \textbf{Acknowledgements}: The authors wish to thank the laboratory Cyceron, Institut Blood and Brain (Normandie Universit\'e, INSERM) where the sheep experiments were initially performed. In particular, we would like to thank Mikael Naveau, Palma Pro Sistiaga, Romaric Saulnier, Cyrille Orset and Denis Vivien. We would also like to thank Christine Chappard for the micro-CT imaging.\\

\noindent \textbf{Funding}:  This project has received funding from the European Research Council (ERC) under the European Union's Horizon 2020 research and innovation program (grant agreement nos. 772786 and 610110, ResolveStroke and REMINISCENCE projects, OC and AA, respectively). \\

\noindent \textbf{Author contributions}: \\
OC contributed to the conception of the ULM study.\\
AA initiated and supervised the UMI part of the project. \\
AC, FB and LD contributed to the conception of the ultrasound sequences.\\
LD and AC contributed to data acquisition.\\
FB performed the UMI analysis. \\
LD and AC performed the ULM analysis.\\
LD, FB, AC, OC and AA contributed to the interpretation of the experimental data.\\
FB, LD, AC and AA prepared the figures.\\
FB, LD and AA contributed to drafting the manuscript.\\
FB, LD, AC, MF, OC and AA discussed the results and contributed to finalizing the manuscript.\\

\noindent \textbf{Competing interests}: AA and MF are inventors on a patent related to UMI held by CNRS (no. US11346819B2, published May 2022). AC and OC are inventors on a patent related also held by CNRS (no. US20240041423A1, published August 2024). OC is co-founder and shareholder of the startup ResolveStroke. All authors declare that they have no other competing interests.\\

\noindent \textbf{Data and materials availability}: \alex{The ultrasound data generated in this study are available at Zenodo~\cite{Bureau2025} (\href{https://zenodo.org/records/15148477}{https://zenodo.org/records/15148477}).}\\

\clearpage

\renewcommand{\thetable}{S\arabic{table}}
\renewcommand{\thefigure}{S\arabic{figure}}
\renewcommand{\theequation}{S\arabic{equation}}
\renewcommand{\thesection}{S\arabic{section}}
\renewcommand{\thesubsection}{S\arabic{section}.\arabic{subsection}}

\setcounter{equation}{0}
\setcounter{table}{0}

\setcounter{page}{1}

\newpage

\counterwithin*{figure}{part}
\counterwithin*{table}{part}

\stepcounter{part}

\noindent{\huge Supplementary Material}

\section{Virtual source basis}
\label{virtual}
\begin{figure}[htbp]
\includegraphics[width=\textwidth]{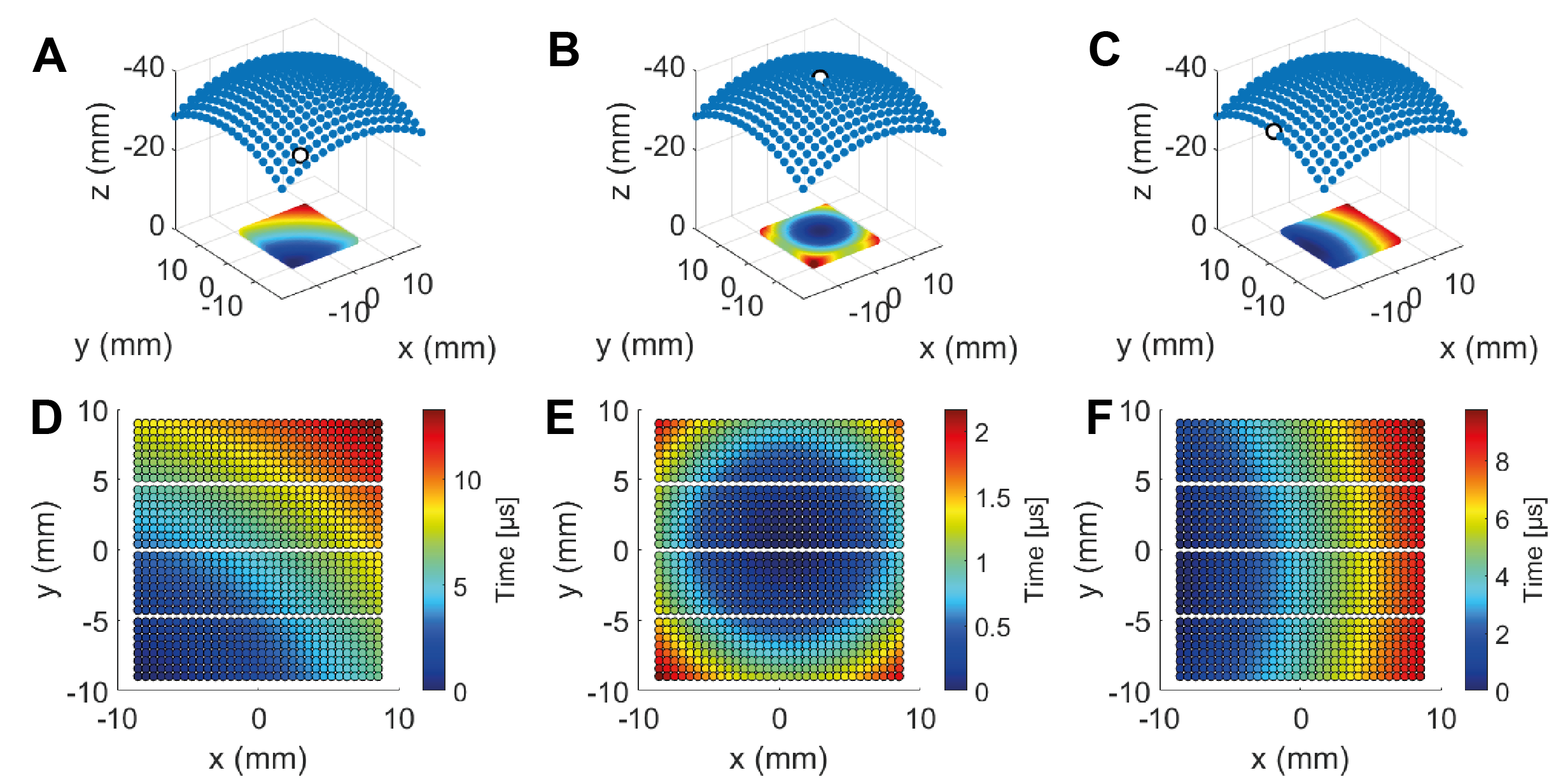}
    \caption{\textbf{Virtual source basis.} (\textbf{A},\textbf{B},\textbf{C}) The $\R-$matrix is acquired with a set of spherical diverging wave associated with virtual sources whose spatial distribution is displayed with blue dots. Each panel corresponds to a different virtual source whose position is indicated by a white disk. (\textbf{D},\textbf{E},\textbf{F}) Time delay applied to transducers to generate the diverging wave associated with the virtual source highlighted by a white disk in panels {A}, B and C, respectively. }
    \label{fig_supp6_virtual_sources}
\end{figure}

\begin{figure}[htbp]
\includegraphics[width=\textwidth]{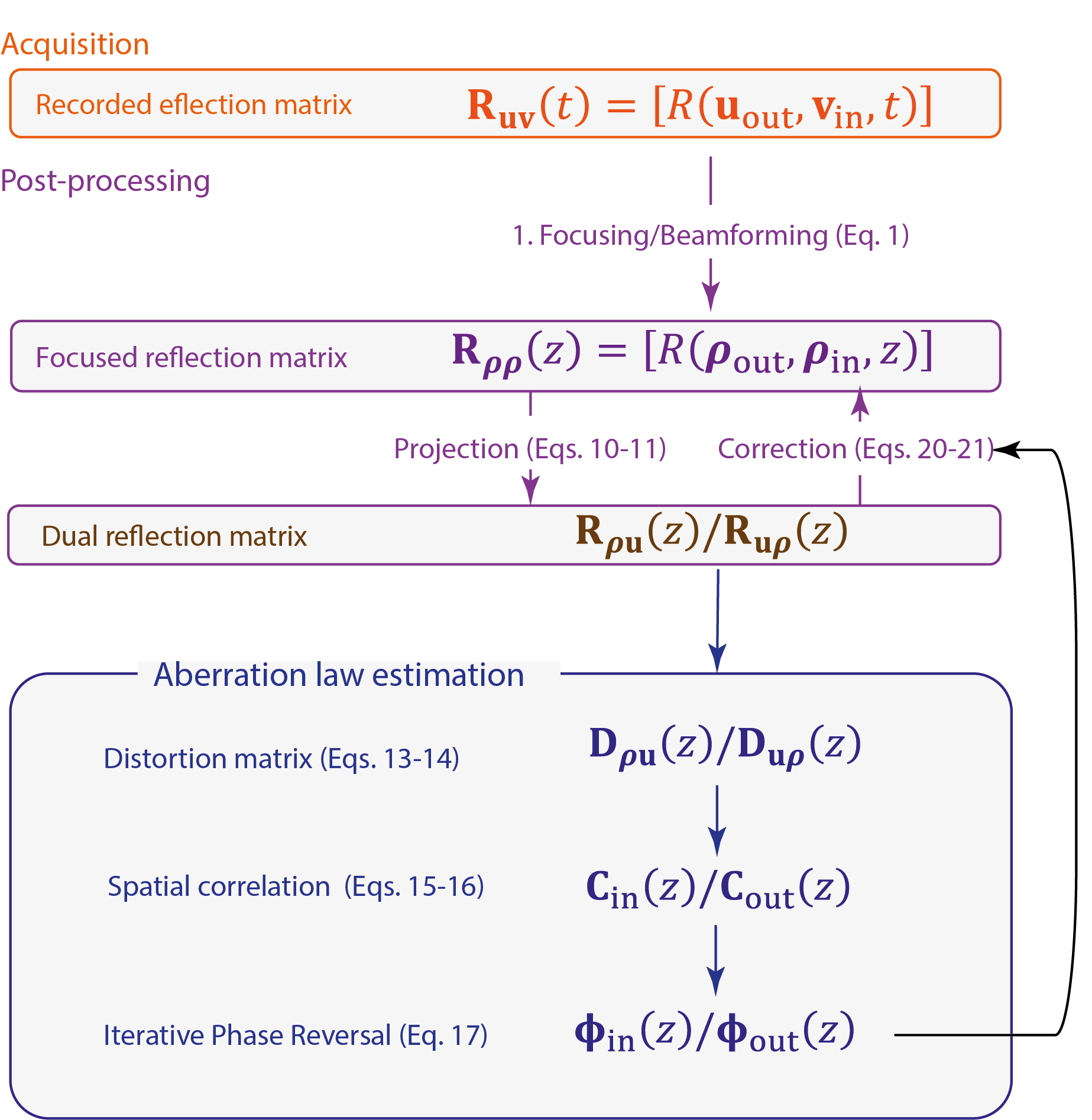}
    \caption{\textbf{Flow chart of the UMI process.}}
    \label{fig_flowchart}
\end{figure}

The distribution of virtual sources considered for the acquisition of the reflection matrix (UMI acquisition) is displayed in Fig.~\ref{fig_supp6_virtual_sources}a. As described in the Methods Section, the virtual sources are located on a sphere centered on the probe midpoint. The time delays applied to each transducer to create the virtual source associated with the white spot in Fig.~\ref{fig_supp6_virtual_sources}A are shown in Fig. \ref{fig_supp6_virtual_sources}B.

\section{Flow chart of the UMI process}
\label{flowchart}

\newpage

\section{Magnitude of aberration correction}
\label{strehl_section}

\begin{figure}[htbp]
\includegraphics[width=\textwidth]{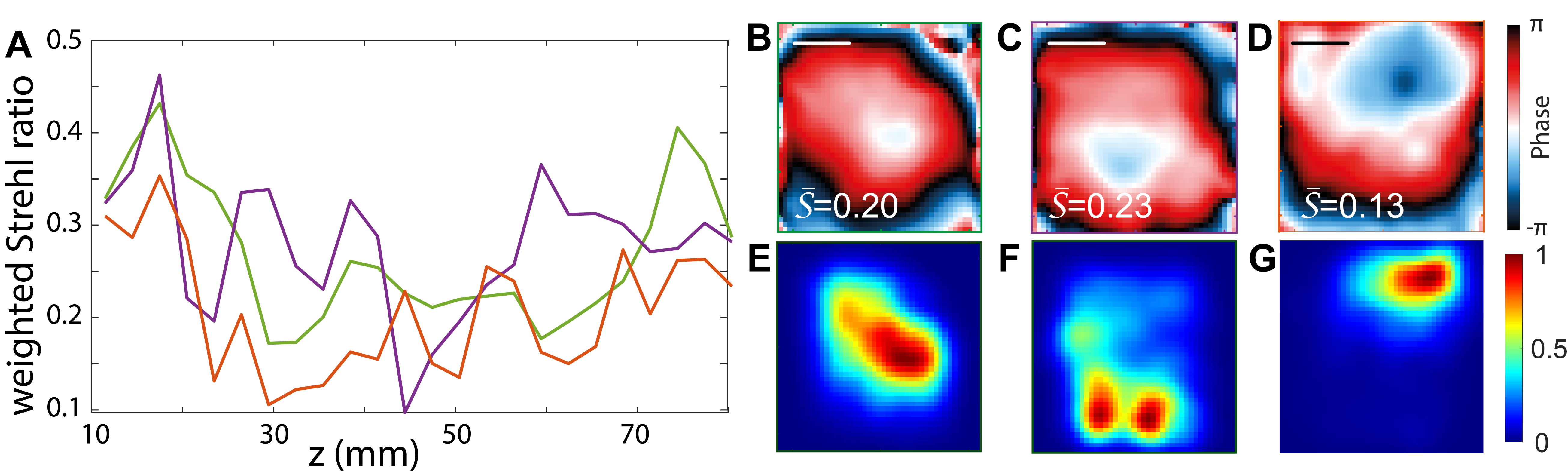}   
\caption{\alex{\textbf{Quantifying the magnitude of aberration correction.} (\textbf{A}) Depth-evolution of the weighted Strehl ratio (Eq.~\ref{Strehl_weight}) associated with the aberration transfer functions extracted for sheep n$^{\textrm{o}}$4 (green), n$^{\textrm{o}}$5 (purple) and n$^{\textrm{o}}$6 (orange). (\textbf{B},\textbf{C},\textbf{D}) Aberration phase laws extracted at depth $z=35$ mm for sheeps n$^{\textrm{o}}$ 4, 5 and 6, respectively (scale bar: 5 mm). (\textbf{E},\textbf{F},\textbf{G}) Modulus of the aberration transfer function (Eq.~\ref{mod}) estimated  at depth $z=35$ mm for sheeps n$^{\textrm{o}}$ 4, 5 and 6, respectively.}}
    \label{fig_supp_strehl}
\end{figure}

\alex{The impact of skull aberrations can be estimated by the Strehl ratio which directly quantifies the decrease of the PSF. \alex{Under a random phase screen model, this Strehl ratio can be shown to scale as $\exp(-\sigma_{\phi}^2)$, with $\sigma_{\phi}$, the standard deviation of the aberration phase law.} However, in the present case, the skull transfer function $\mathbf{A}=[A(\mathbf{u},z)]$ is not of constant amplitude. It can actually be estimated through the following matrix product:
   \begin{equation}
   \label{mod}
   \bm{A}(z)= \mathbf{C} \times \exp \left ( i \bm{\Phi}(z) \right) .
   \end{equation}
Examples of phase and modulus of the estimated transfer function $\bm{A}(z)$ are displayed in Figs.~\ref{fig_supp_strehl}B and C, respectively. While the phase of $\bm{A}(z)$ is by definition equal to the aberration phase law $\bm{\Phi}(z)$, its modulus describes the wave amplitude across the probe aperture. As shown by Fig.~\ref{fig_supp_strehl}, this amplitude is clearly not constant over the probe aperture. The evaluation of aberration correction magnitude should therefore include this amplitude variation of the aberration transfer function. To do so, a weighted Strehl ratio can be defined in order to account for both the amplitude and phase of the aberration transfer function:
\begin{equation}
\label{Strehl_weight}
\alex{\bar{\mathcal{S}}(z)=\frac{\left |\langle {A}(\mathbf{u},z)\phi(\mathbf{u},z) \rangle_{\mathbf{u}}
\right |^2}{\langle |{A}(\mathbf{u},z){\phi}(\mathbf{u},z)|^2 \rangle_{\mathbf{u}}}.}
\end{equation}
This normalized Strehl ratio is an important parameter for ULM since it corresponds the correlation degree $\gamma$ between the imaging PSF and its theoretical diffraction-limited value.}

\alex{The depth-evolution of $\bar{\mathcal{S}}(z)$ is displayed for the three sheeps in Fig.~\ref{fig_supp_strehl}A. The weighted Strehl ratio is lower for sheep n$^{\textrm{o}}$6 $\left (\langle \bar{S} \rangle \sim 0.21 \right )$ than for sheeps n$^{\textrm{o}}$4 and 5 $\left (\langle \bar{S} \rangle \sim 0.27 \right .$ and 0.28, respectively). This lower Strehl ratio indicates a stronger aberration magnitude for sheep n$^{\textrm{o}}$6. This observation is in agreement with the stronger distortions exhibited by the initial RPSF for sheep n$^{\textrm{o}}$6 (Fig.~\ref{fig7_skull_microct}B) and is explained by the complex structure of its skull (Fig.~\ref{fig7_skull_microct}A).}

\section{Resolution enhancement}
\label{resenh}
\begin{figure}[htbp]
\includegraphics[width=\textwidth]{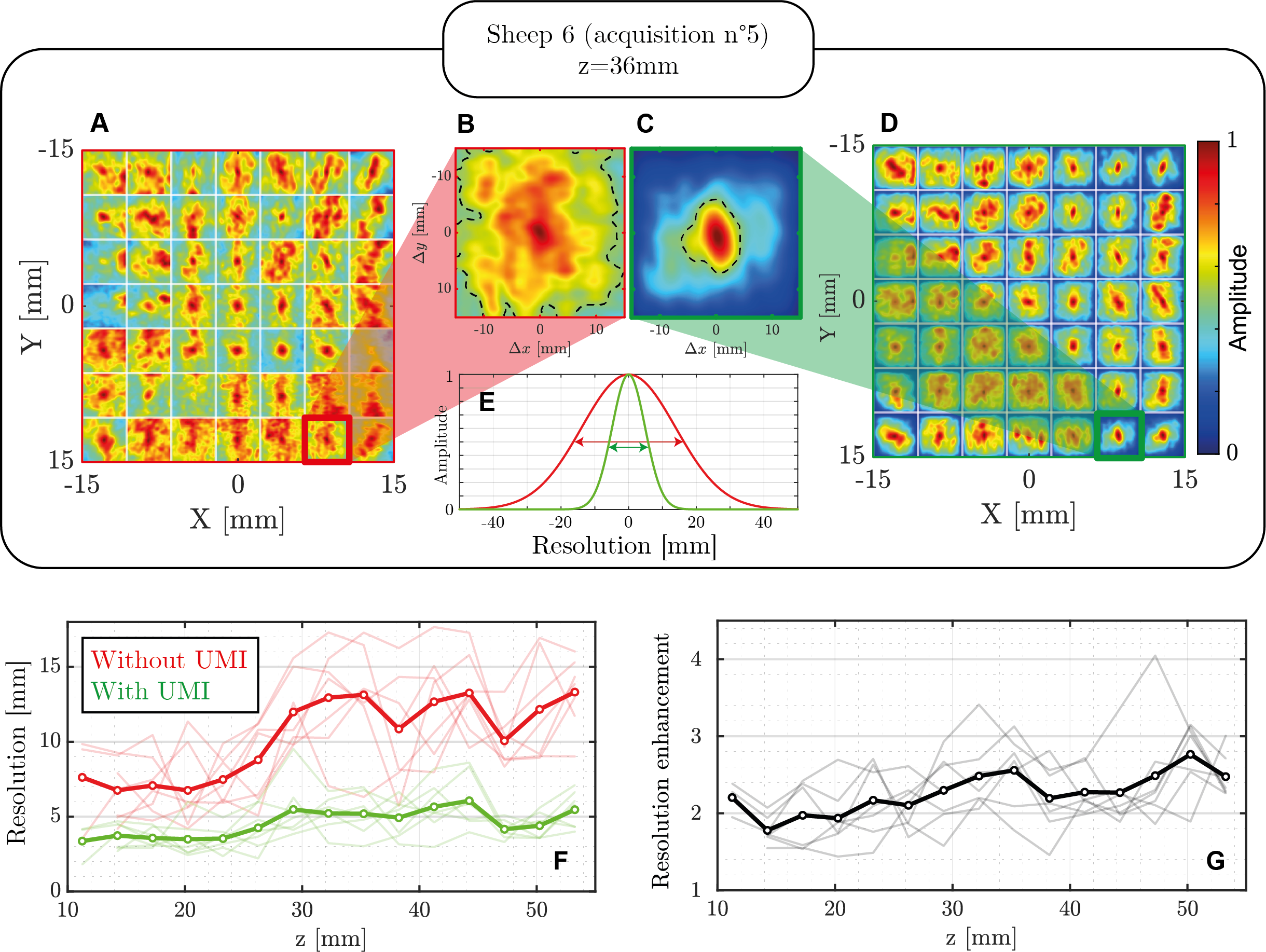}   \caption{\textbf{Quantifying the ultrasound image resolution.} (\textbf{A},\textbf{D}) RPSF map at $z=36$ mm before and after correction, respectively. (\textbf{B},\textbf{C}) Example of one RPSF at a specific location with the -3 dB-area surrounded by a dashed contour, before and after correction, respectively. (\textbf{E}) Radial average of the RPSFs displayed in panels C (red line) and D (green line). (\textbf{F}) Resolution $\delta \rho $ as a function of depth before (red) and after UMI correction (green). The transparent curves refer to the seven individual data sets, while the opaque curves represent their average. (\textbf{G}) Resolution enhancement provided by UMI. The gray lines refer to each individual data set, the black line shows their average value.}
    \label{fig_supp3_resolution}
\end{figure}

The method used for the resolution estimation is illustrated in Fig. \ref{fig_supp3_resolution} by considering an example of RPSF for a specific data set (sheep 6, acquisition n$^{\textrm{o}}$5) and at a specific depth ($z=36$ mm). The RPSF map before and after the correction is shown in Figs.~\ref{fig_supp3_resolution}A,D. An example of RPSF is displayed in Fig. \ref{fig_supp3_resolution}B,C. The -3 dB-area, noted $\mathcal{A}$, is evaluated for each RPSF. An estimate of the resolution length $\delta\rho$ is then deduced using Eq.~\ref{resolution} (Fig. \ref{fig_supp3_resolution}E).
The depth evolution of the resolution $\delta \rho$ is shown for each data set in Fig. \ref{fig_supp3_resolution}F. The benefit of UMI in terms of resolution is clear whatever the acquisition. The gain in resolution displayed in Fig. \ref{fig_supp3_resolution}G increases from a factor 2 at shallow depths to 3 at large depths.

\clearpage 

\section{Multiple scattering rate}
\label{msr}
\begin{figure}[htbp]
\includegraphics[scale=0.5]{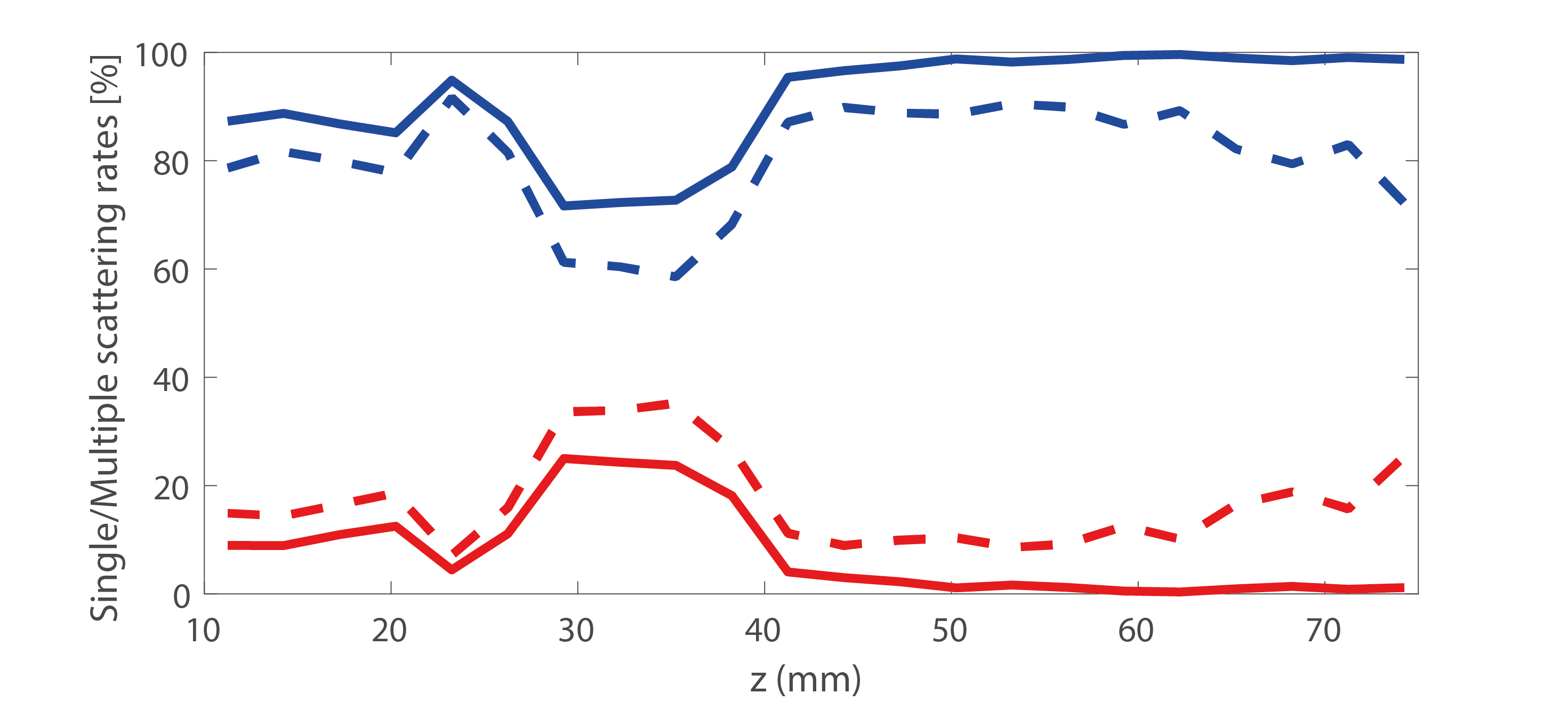}
    \caption{\textbf{Single and multiple scattering rates.} (\textbf{A}) Depth evolution of the single scattering (blue) and multiple scattering (red) rates without (dashed line) and with UMI (continuous line). (\textbf{B}) Ratio between the multiple scattering rate measured after and before UMI. The y-scale is in dB. The ultrasound data shown here corresponds to sheep n$^{\textrm{o}}$6, acquisition 4.}
    \label{figMS}
\end{figure}

Figure~\ref{figMS}A displays the depth evolution of the single and multiple scattering rates (Methods) before and after UMI for the acquisition 4 on sheep n$^{\textrm{o}}$6. Corresponding RPSF maps and confocal volumes are displayed in Fig.~\ref{fig2_UMI_analysis}. The expression of the multiple scattering rate $\alpha_M$ is provided in the Methods (Eq.~\ref{multiple}) and the single scattering rate $\alpha_S$ is deduced from $\alpha_M$ as follows: $\alpha_M=1-\alpha_S$. The reduction of the multiple scattering rate provided by UMI is displayed in Fig.~\ref{figMS}B.

\clearpage 

\section{\alex{Comparison \flav{with} other aberration correction methods}}
\label{other}
\alex{The UMI framework offers the opportunity to compare our approach with other aberration correction methods. In particular, our approach shares some similarities with the iterative time-reversal (ITR) approach of speckle noise developed by Montaldo et al.~\cite{montaldo_time_2011}. While matrix imaging is inspired by this work and others studies in the field of ultrasound imaging ~\cite{varslot_eigenfunction_2004,robert_greens_2008}, it features several distinct and important advances. The first is its primary building block: the broadband focused reflection matrix that precisely selects the echoes originating from a single scattering event at each depth. This \flav{time-gating} operation is decisive in terms of signal-to-noise ratio since it drastically reduces the detrimental contribution of out-of-focus and multiply scattered echoes. Second, the IPR process used in the matrix imaging process is more efficient than an iterative time reversal process. Time reversal is an adaptive filter~\cite{tanter_time_2000} that is far from being optimal for imaging since it acts as a low pass spatial frequency filter.} 

\begin{figure}[htbp]
\includegraphics[width=\textwidth]{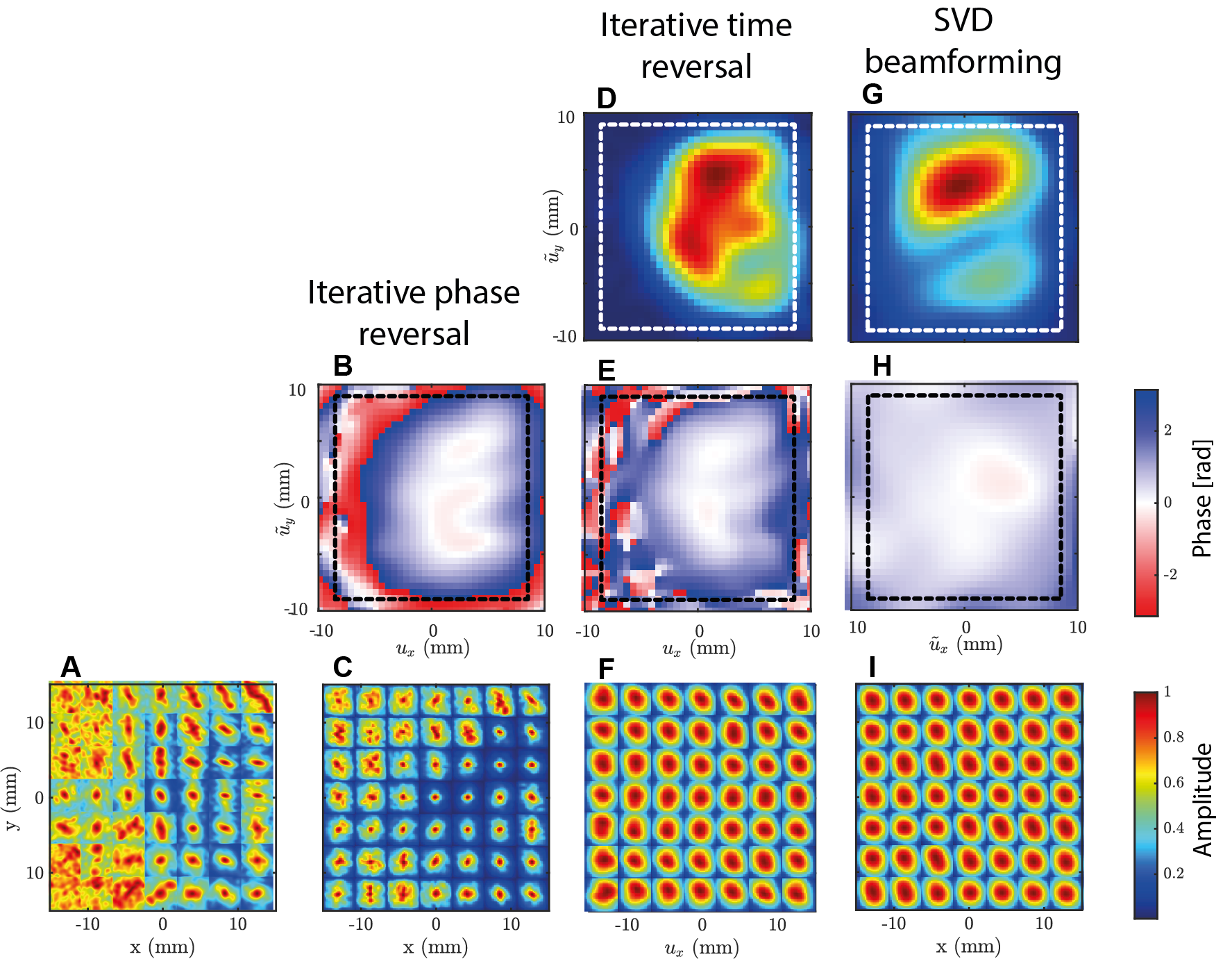}
    \caption{\alex{\textbf{Comparison with other aberration correction methods.} (\textbf{A}) RPSF measured at depth $z=56$ mm in sheep n$^{\textrm{o}}$6. (\textbf{B}) Corresponding aberration phase law extracted by an iterative phase reversal process (Eq.\ref{recursive}). (\textbf{C}) RPSFs after iterative phase reversal processing (Eqs.~\ref{T}, \ref{cor1} and \ref{cor2}).  (\textbf{D}) Modulus of the time-reversal invariant (Eq.~\ref{T2}) extracted in the transducer basis. (\textbf{E}) Phase of the time-reversal invariant $\mathbf{U}$ (Eq.~\ref{T2}) extracted in the transducer basis. (\textbf{F}) RPSFs after iterative time reversal reversal processing (Eq.~\ref{recursive2}) operated in the transducer basis. (\textbf{G}) Modulus of the first eigenvector $\mathbf{V}$ extracted by SVD of the distortion matrix expressed in the virtual source basis. (\textbf{H}) Phase of the first eigenvector extracted by SVD in the virtual source basis. (\textbf{I}) RPSFs after SVD beamforming (Eq.~\ref{T3}).}}
    \label{figCorrComparison}
\end{figure}

\alex{This detrimental effect can be highlighted by implementing the ITR approach in the UMI pipeline. To do so, the IPR process (Eq.~\ref{recursive}) should be replaced by ITR, such that:
   \begin{equation}
   \label{recursive2}
  \sigma \mathbf{U}_{\inm/\outm}(z)=\mathbf{C}_{\textrm{in/out}} \times \mathbf{U}_{\textrm{in/out} }(z).
   \end{equation}
The time-reversal invariant $\mathbf{U}_{\inm/\outm}$ corresponds to the first eigenvector of the correlation matrix $\mathbf{C}_{\textrm{in/out}}$ and $\sigma$ is the associated eigenvalue. The time-reversal invariant yields an estimator of the input/output transmission matrix between the probe and each focal plane:
\begin{equation}
\label{T2}
\mathbf{T}_{\textrm{in/out}}(z)=\mathbf{U}_{\textrm{in/out} }(z) \circ \mathbf{T}_0 
\end{equation}
The result of the ITR process is illustrated in Fig.~\ref{figCorrComparison} by applying it to the focused reflection matrix at depth $z=56$ mm in sheep n$^{\textrm{o}}$6, acquisition 4. Even if the phase of the time-reversal invariant (Fig.~\ref{figCorrComparison}E) shares some similarities with the phase law resulting from the IPR process (Fig.~\ref{figCorrComparison}B), it also exhibits random fluctuations on some parts of the probe . Those areas coincide with the low amplitude displayed the time-reversal invariant (Fig.~\ref{figCorrComparison}D), especially on the left of the probe. This strong variation in amplitude is a manifestation of the angular filtering process induced by iterative time reversal on the received signals. When the focused reflection matrix is beamformed (Eqs.~\ref{cor1} and \ref{cor2}) using the time reversal invariant (Eq.~\ref{T2}), this angular filtering of ultrasound data results in enlarged focal spots, as illustrated by the resulting map of RPSFs shown in Fig.~\ref{figCorrComparison}F. The comparison with the IPR process (Fig.~\ref{figCorrComparison}C)  therefore demonstrates the benefit of our approach compared to an ITR approach  [54]. While the IPR process narrows the confocal component of each RPSF and lowers the multiple scattering background (Fig.~\ref{figCorrComparison}C), ITR acts as an adaptive filter and enlarges each focal spot. This loss of resolution would be detrimental to ULM since it would affect the SNR and the bubble localization precision.}
\newpage

\alex{The same observation holds for the SVD beamformer proposed by Bendjador et al.~\cite{bendjador_svd_2020}. This approach is mathematically equivalent to iterative time reversal but operated from the insonification basis, i.e here the virtual source basis~\cite{Bendjador2021}. This basis and, equivalently, the plane wave basis are actually not adequate since most of aberrations are induced by the skull heterogeneities, \textit{i.e} close to the probe. Again, we can show it by implementing the SVD beamformer in the UMI framework. To do so, a distortion matrix \flav{$\mathbf{D}_{\mathbf{v}\bm{\rho}}$} can be built in the virtual source basis, following the same recipe as in the transducer basis (Eq.~\ref{D2}), and its first singular vector $\mathbf{V}_{\textrm{in}}$ directly provides an estimator of the transmission matrix $\mathbf{T}_{\textrm{in}}$, such that:
\begin{equation}
\label{T3}
\mathbf{T}_{\textrm{in}}(z)=\mathbf{V}_{\textrm{in} }(z) \circ \mathbf{T}_0 
\end{equation}
Because aberrations are from being isoplanatic from the virtual source basis, the aberration phase law exhibited by $\mathbf{V}_{\textrm{in}}$ is almost flat (Fig.~\ref{figCorrComparison}H), resulting in an extremely limited aberration correction effect. Moreover, as for the ITR approach, the SVD provides an eigenvector $\mathbf{V}_{\textrm{in}}$ that shows fluctuations in amplitude. It therefore acts as an angular filter on the corrected wave-field (Fig.~\ref{figCorrComparison}G). As for ITR (Fig.~\ref{figCorrComparison}F), the SVD beamformer results in enlarged focal spots (Figure~\ref{figCorrComparison}I). The SVD beamformer would therefore be inefficient for improving the ULM process. Note that an even worse result would be obtained when the SVD beamformer is operated from the plane wave basis. }

\clearpage
\section{Quantitative comparison between MRA and ULM images}
\label{comparison}

\begin{figure}[htbp]
\includegraphics[width=\textwidth]{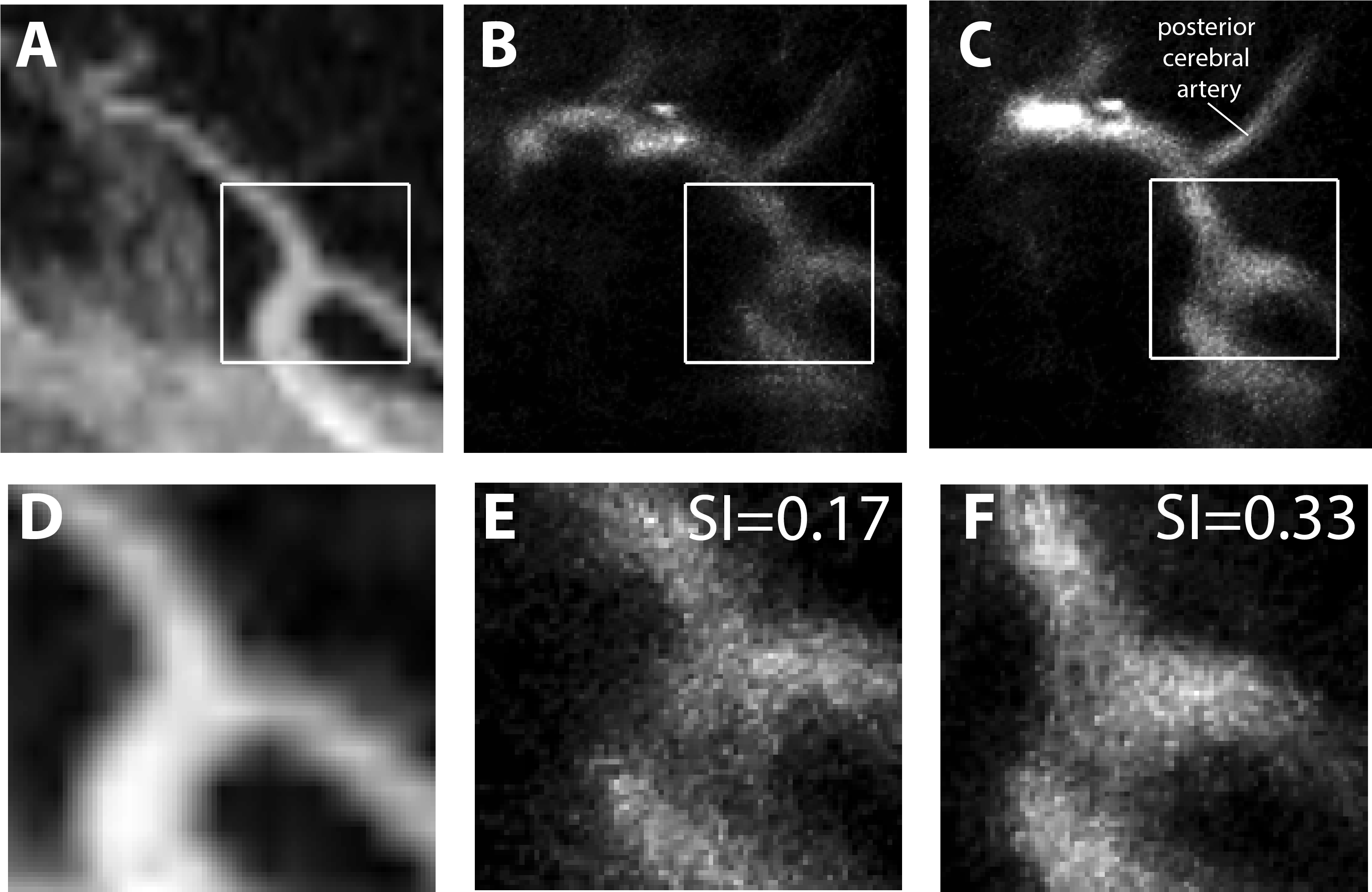}
    \caption{\alex{\textbf{Quantitative comparison between ULM and MRA cross-sections in  the saggital plane.} (\textbf{A}) MRA. (\textbf{B}) Raw ULM image. (\textbf{C}) ULM image after UMI correction. (\textbf{D},\textbf{E},\textbf{F}) Zoom on the junction between middle cerebral and posterior communicating arteries displayed as a white rectangle in panels A, B and C, respectively. For panels E and F, a structural similarity index (SI) is computed between the corresponding ULM image and the MRA image displayed in panel D.}}
    \label{figComparison0}
\end{figure}

\alex{For a quantitative comparison between MRA and ULM images, we will consider an inset at the junction of the posterial cerebral arteries and the Willis polygon already visible on the saggital slice displayed in Fig.~\ref{fig6_mra}B and D. This part of the vascular network is nicely resolved by MRA (Fig.~\ref{figComparison0}D). It can therefore be used here as a ground truth for corresponding ULM images displayed in Fig.~\ref{figComparison0}E and F. The effect of UMI can be quantitatively evaluated by considering the structural similarity index (\flav{SI}) of each ULM image with MRA: \flav{SI}=0.17 for raw ULM vs. \flav{SI}=0.33 with UMI. The SI index is therefore  multiplied by a factor 2 in this specific part of the image.}

\begin{figure}[htbp]
\includegraphics[width=\textwidth]{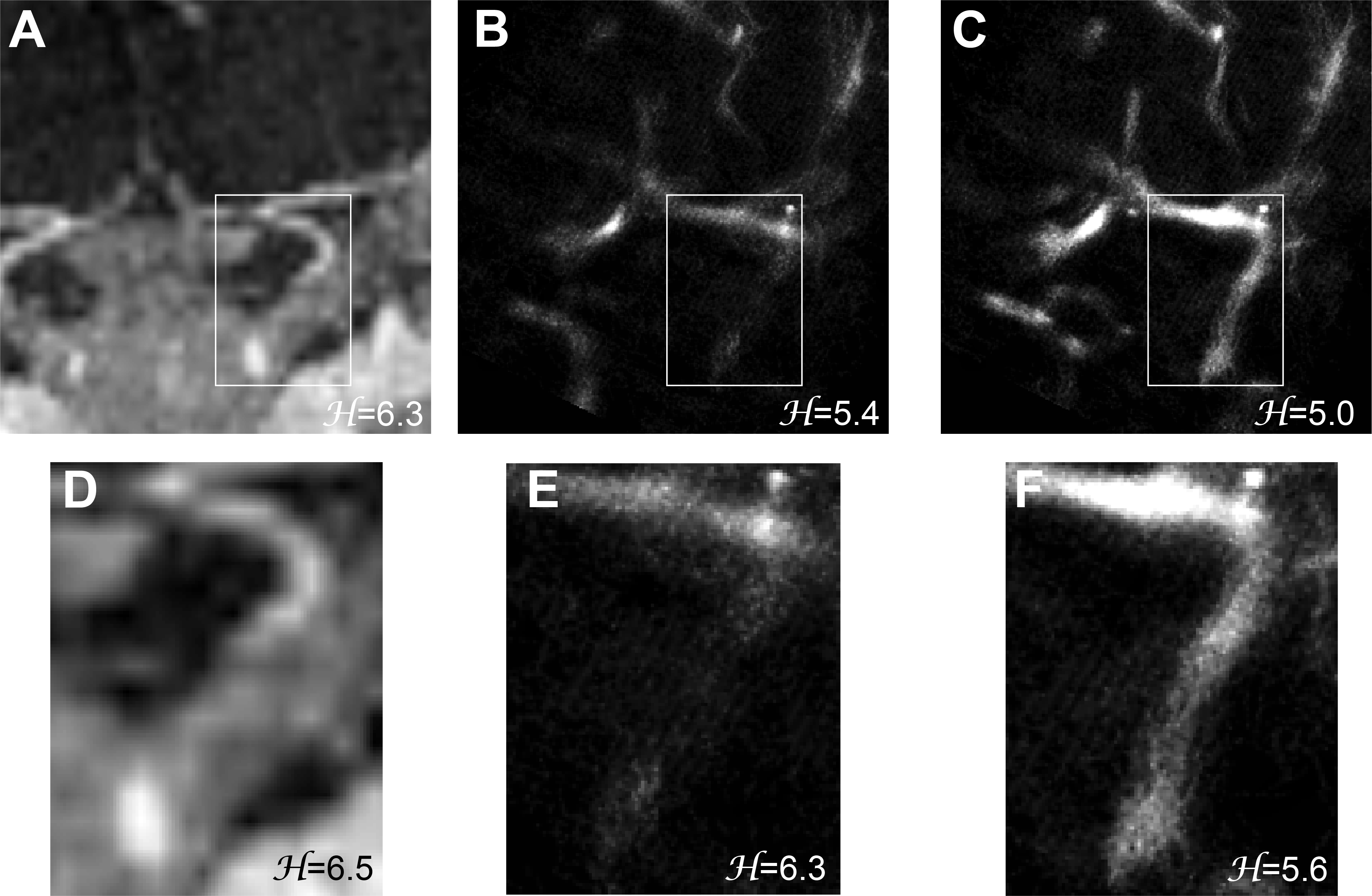}
    \caption{\alex{\textbf{Quantitative comparison between ULM and MRA cross-sections in the coronal plane.} (\textbf{A}) MRA. (\textbf{B}) ULM. (\textbf{C}) ULM combined with UMI. (\textbf{D},\textbf{E},\textbf{F}) Zoom on the right branch of the Willis circle depicted by the white rectangle in panels A, B and C, respectively. The entropy $\mathcal{H}$ of each image is indicated on each panel.}}
    \label{figComparison}
\end{figure}

\alex{ULM can do better than MRA by revealing the presence of smaller vessels~\cite{Denis2025}. This higher resolving power of ULM is reinforced by UMI, both in: (\textit{i}) the coronal plane with the emergence of the posterior cerebral artery in Fig.~\ref{figComparison0}C that was not visible in the MRA image Fig.~\ref{figComparison0}A; (\textit{ii}) the sagittal plane with the resolution of smaller vessels on top of the Willis polygon in Fig.~\ref{figComparison}C that are not revealed by MRA (Fig.~\ref{figComparison}A). The latter cross-section highlights the drastic gain in contrast and resolution provided by the combination of ULM and UMI (Figs.~\ref{figComparison}C and F) compared to gold standard MRA (Figs.~\ref{figComparison}A and D) and raw ULM (Figs.~\ref{figComparison}B and E). This improvement cannot be here quantified with a structural similarity index since the MRA image is of too low quality to be used as a ground truth. However, the image quality can be quantified by its entropy $\mathcal{H}$~\cite{Tsai2007} that measures its degree of randomness. The value of  $\mathcal{H}$ is reported on each image displayed in Fig.~\ref{figComparison}. While ULM only leads to a slight decrease of $\mathcal{H}$ by 0.2 compared to MRA (Figs.~\ref{figComparison}D,E), the entropy loss is increased by almost a factor 5 with UMI (Fig.~\ref{figComparison}F).} 

\clearpage
\section{ULM images for each dataset}
\label{ulmall}
\begin{figure}[htbp]
\includegraphics[scale=0.75]{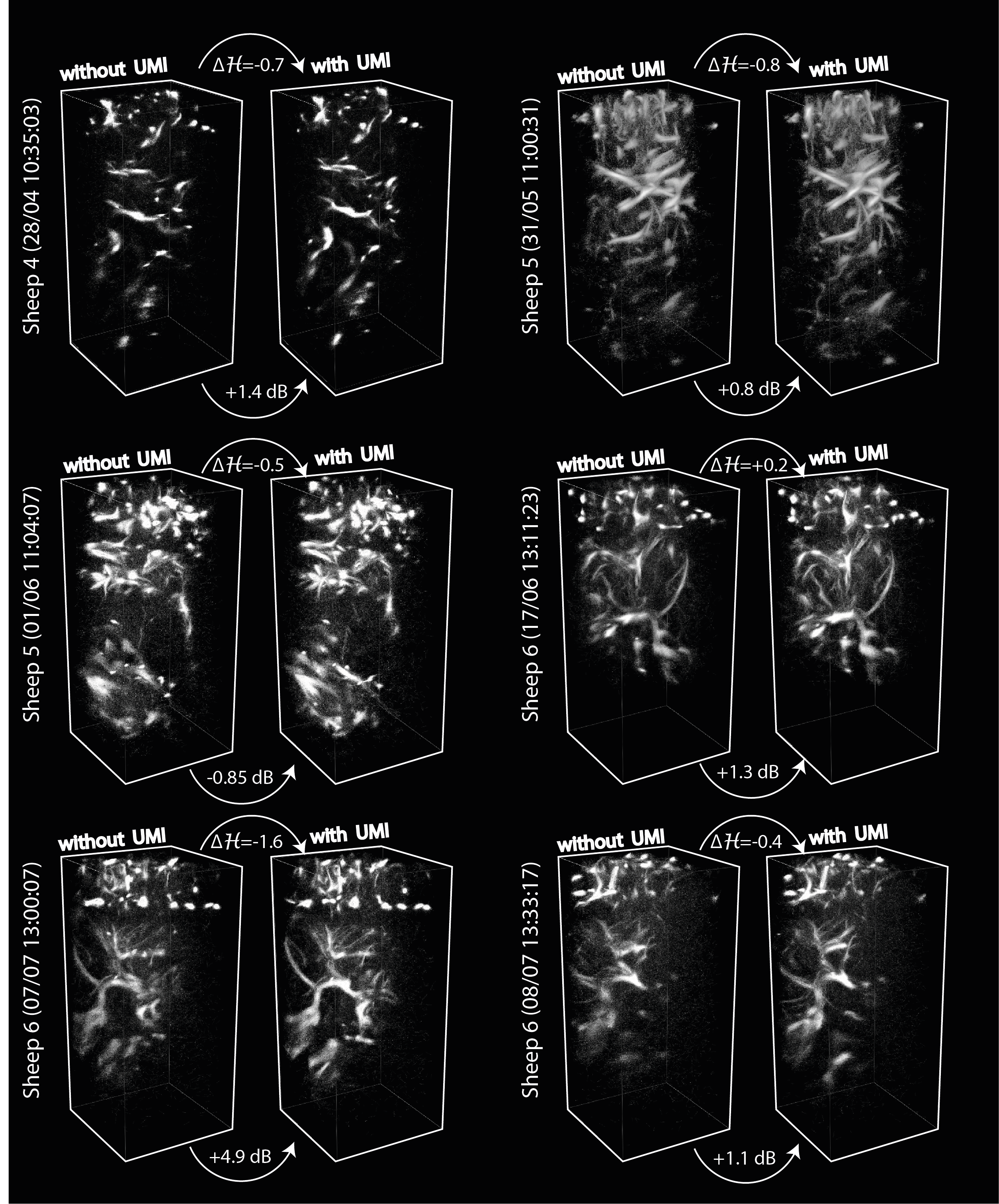}
    \caption{\textbf{Impact of UMI on different ULM  acquisitions.} ULM volumes with (left) and without (right) application of the UMI correction. Several data acquisitions on three different sheeps are presented here (see also Movie S2).}
    \label{fig_supp1_ulm_volume}
\end{figure}

Figure~\ref{fig_supp1_ulm_volume} compares the ULM images obtained with and without UMI for different acquisitions on the three sheeps (Tab.~\ref{tabAcq}). In all cases, a clear resolution and contrast improvement is observed for the different viewing angle (i.e. different probe position) in each acquisition. To better visualize such data in 3D, corresponding films in which the viewing angle changes are available in Movie S2. 

\alex{The image improvement can be quantified by the loss of entropy $\Delta \mathcal{H}$ and the contrast improvement after UMI correction. Both quantities are indicated in each image of Fig.~\ref{fig_supp1_ulm_volume} for depths ranging from 20 to 50 mm. The contrast is defined as the ratio between the variance and the mean intensity of each ULM image. Those quantities are averaged over the pixels of each ULM image for which bubbles have been detected. Zero-value pixels are discarded since they would bias the estimation of entropy and contrast.} 

\clearpage

\begin{table}[ht]
\normalsize
{%
\begin{tabular}{|c|c|c|c|c|c|c|c|c|ll}
\cline{4-7}
\multicolumn{3}{c|}{} & \multicolumn{1}{c|}{\textbf{UMI}}  & \multicolumn{3}{c|}{\textbf{ULM}} &  \multicolumn{2}{c}{} & &  \\ 
\cline{1-8}
\textbf{\begin{tabular}[c]{@{}c@{}}Sheep \\ number\end{tabular}} & \textbf{\begin{tabular}[c]{@{}c@{}}Acq.\\ number\end{tabular}} & \begin{tabular}[c]{@{}c@{}}\textbf{Date} \\ (day/month)\end{tabular} & \textbf{\begin{tabular}[c]{@{}c@{}}Time \end{tabular}} & 
\textbf{\begin{tabular}[c]{@{}c@{}}Time\end{tabular}} & \begin{tabular}[c]{@{}c@{}}\textbf{Injection} \\ \textbf{volume} \\ (mL/min)\end{tabular} & \textbf{\begin{tabular}[c]{@{}c@{}}Nb of \\ injections\end{tabular}} & \textbf{\begin{tabular}[c]{@{}c@{}}Figures \end{tabular}}     \\ \cline{1-8}
4 & 1 & 28/04 & 11:27 & 10:35 & 2 & 6 & \ref{fig7_skull_microct},\ref{fig_supp_strehl},\ref{fig_supp3_resolution},\ref{fig_supp1_ulm_volume}  \\ \cline{1-8}
\multirow{2}{*}{S5}       
& 1 & 31/05 & 11:21 & 11:00 & 2 & 3 &  \ref{fig_supp3_resolution}, \ref{fig_supp1_ulm_volume}  \\ 

& {2} & {01/06} & {11:43} & {11:04} & {1.5} & {6} & \ref{fig7_skull_microct}, \ref{fig_supp_strehl}, \ref{fig_supp3_resolution},\ref{fig_supp1_ulm_volume}   \\ 
\cline{1-8}
\multirow{6}{*}{S6}                                                       & 1 & 17/06 & 10:33 & 10:40 & 1.5 & 5 &  \ref{fig_supp3_resolution}    \\ 

& 2 & 17/06 & 10:33 & 10:58 & 2 & 5 &\ref{fig_supp3_resolution}     \\
& 3 & 17/06 & 13:33 & 13:11 & 2.5 & 5 &  \ref{fig_supp1_ulm_volume}      \\ 

& {4} & {07/07} & {11:49} & {13:00} & {1.75} & {7} & \ref{fig2_UMI_analysis},\ref{fig3_ULMvolume},\ref{fig4_localization},\ref{fig6_mra}, \ref{fig_supp_strehl},\ref{fig_supp3_resolution}, \ref{figMS},\ref{figCorrComparison},\ref{fig_supp1_ulm_volume}\\ 

& {5} & {08/07} & {12:36} & {13:33} & {1.5} & {10} & \ref{fig7_skull_microct},\ref{fig_supp3_resolution}, \ref{fig_supp1_ulm_volume}    \\ 
\cline{1-8}
\end{tabular}%
}
\caption{\textbf{Experimental conditions for each acquisition.} }
\label{tabAcq}
\end{table}

\clearpage
\section*{Captions of Supplementary Movies}

\noindent \alex{Movie S1: {\textbf{Result of UMI at each depth}. Left: Selected depth. Middle Top: Local RPSFs before and after correction. Middle Bottom: Averaged RPSF before and after correction. Top right: Extracted aberration phase law. Bottom right: Resolution enhancement provided by UMI. The data shown here correspond to sheep n$^{\textrm{o}}$6 (acquisition 4)}} \\

\noindent \alex{Movie S2: {\textbf{ULM images for different acquisitions before and after UMI correction}. 3D images are shown as maximum intensity projections.}} \\

\end{document}